\documentclass[aps,prb]{revtex4}
\usepackage{graphicx,amssymb,amsfonts,amsmath}

\begin{document}

\title{Dissipative and nonequilibrium effects near a superconductor-metal
quantum critical point}
\author{Aditi Mitra}
\affiliation{Department of Physics, New York University, 4
Washington Place, New York, NY 10003}
\date{\today}

\begin{abstract}
We present a microscopic derivation of the effect of current flow on
a system near a superconductor-metal quantum critical point. The
model studied is a $2d$ itinerant electron system where the
electrons interact via an attractive interaction and are coupled to
an underlying normal metal substrate which provides a source of
dissipation, and also provides a source of inelastic scattering that
allows a nonequilibrium steady state to reach. A nonequilibrium
Keldysh action for the superconducting fluctuations on the normal
side is derived. Current flow, besides its minimal coupling to the
order parameter is found to give rise to two new effects. One is a
source of noise that acts as an effective temperature $T_{eff} = e E
v_F \tau_{sc}$ where $E$ is the external electric field, $v_F$ the
Fermi velocity, and $\tau_{sc}$ is the escape time into the normal
metal substrate. Secondly current flow also produces a drift of the
order-parameter. Scaling equations for the superconducting gap and
the current are derived and are found to be consistent with previous
phenomenological treatments as long as a temperature $T \sim
T_{eff}$ is included. The current induced drift is found to produce
additional corrections to the scaling which are smaller by a factor
of ${\cal O}(\frac{1}{E_F \tau_{sc}})$, $E_F$ being the Fermi
energy.
\end{abstract}

\maketitle

\section{Introduction}

Quantum critical phenomena is the study of how
a system loses long range order at $T \rightarrow 0$
as a parameter of the Hamiltonian is changed~\cite{Sondhi97,Sachdev99}.
The non-commutativity
of position and momentum in quantum mechanics implies that the spatial and temporal
fluctuations of the order
parameter are
coupled to each other at the zero temperature quantum critical point.
The effect of temperature on a quantum critical point  has some generic
features~\cite{Hertz76,Millis93} such as,
a non-zero temperature produces dephasing or decoherence that
cuts off divergences in correlation lengths and times. Thermal decoherence also decouples
spatial and temporal fluctuations causing
a crossover from quantum to classical behavior.

While quantum phase transitions for systems in equilibrium have been
extensively studied, a much less understood issue is the effect of a
nonequilibrium probe such as current flow on a system in the
vicinity of a quantum critical point. Scaling theories exist which
assume that the primary effect of a nonequilibrium probe is to
produce decoherence or an effective temperature. Thus nonequilibrium
scaling relations are obtained by replacing temperature in the
equilibrium scaling relations by the appropriate nonequilibrium
energy scale~\cite{Sondhi97}. A microscopic treatment to justify
this and in the process also identify the appropriate nonequilibrium
energy scale is often challenging as this requires a treatment that
goes beyond a linear response Kubo formula calculation. Only a
handful of such treatments exist for magnetic-paramagnetic
~\cite{Feldman05,Mitra06,Green06,Mitra08a}, and
superfluid-insulator/metal quantum critical
points~\cite{Phillips04,Green05,TakeiKim07}.

In this paper we revisit the problem of non-linear effects, in
particular the effect of a uniform current flow on a system near a
superconductor-metal quantum critical point. Existing studies have
so far involved writing phenomenological effective theories for a
charged order-parameter in the presence of an electric field and/or
external dissipation~\cite{Phillips04, Green05}. In this paper we
carry out a fully microscopic derivation of the appropriate
nonequilibrium effective theory starting from a fermionic model
under external drive. In doing so, we address the issue of how the
fermionic system reaches a nonequilibrium steady state, and find
that the underlying nonequilibrium fermions give rise to additional
terms in the effective theory for the charged order-parameter that
were previously missed. We then proceed to determine the effect
of these terms on scaling near the equilibrium quantum critical
point. (Note that the observation that nonequilibrium electrons can
significantly modify the scaling near critical points was also
pointed out in~\cite{Sondhi97}, and 
was experimentally observed in thin films of Bi in~\cite{Goldman}).

The geometry that will be studied (shown schematically in
Fig~\ref{schem}) is a 2d itinerant electron system where the
electrons interact with each other via an attractive interaction.
This system is driven out of equilibrium by an in-plane electric
field so that a current flows through the bulk of the system. In
addition, the 2d layer is coupled to an underlying normal metal
substrate with which it can exchange particle as well as energy
which thus serves as a heat sink that allows the layer to reach a
nonequilibrium steady state. The coupling to  the substrate also
provides a source of dissipation that when made sufficiently large
can destroy superconductivity in the layer~\cite{TakeiKim07}. (Note
that the model in Fig~\ref{schem} for the case of repulsive
interactions of the electrons in the layer and for parameters that
are such that the system is near a ferromagnetic-paramagnetic
quantum critical point was studied in~\cite{Mitra08a}).

As we shall show, in equilibrium the effective action for the
superconducting fluctuations on the ordered and the disordered side
has a local Caldiera-Leggett dissipation typical of systems where
particle number is not conserved~\cite{JJunctions}. Effective
theories for superconducting fluctuations with local dissipation
have been extensively studied in equilibrium,
~\cite{Phillips00,Troyer04,Refael07,Chakravarty88}, but hardly at
all out of equilibrium (with the exception of~\cite{Phillips04}).
Out of equilibrium, our microscopic treatment reveals three effects
of current flow, one is the usual minimal coupling of the current to
the charged order-parameter, second is a source of noise which for low
frequencies and long wavelength fluctuations of the order-parameter
essentially acts as an effective temperature which equals the typical 
energy an electron gains on being accelerated by the external electric field.
In our model $T_{eff}= e E v_F
\tau_{sc}$, where $e$ and $v_F$ are the charge and Fermi velocity of
the electrons, and $\tau_{sc}$ is the inelastic scattering
time or escape time into the normal metal reservoirs. Thirdly, we
also find that current flow can cause the order-parameter to drift
with a drift velocity $v_D = \frac{e E}{m}\tau_{sc}$, $m$ being the
mass of the electrons.

We briefly mention the relation of the work presented here to that
of~\cite{Phillips04} which also had an extrinsic dissipation which
was introduced phenomenologically. Thus in their model, current
affects the order-parameter only via its minimal coupling to it, and
the properties of the dissipative reservoir were unaffected by the
current flow. In our model, the dissipation originates via the
coupling of the superconducting order-parameter to the underlying
normal electrons, whose properties are itself modified due to an
external drive. Taking this effect into account shows that the
order-parameter is subjected to a noise and also drifts with the
current. As we shall show, in the quantum-critical regime, current noise 
gives corrections to the scaling which are of ${\cal O}((T_{eff}\tau_{sc})^{2/3})$, where
$T_{eff}\tau_{sc}$  is the ratio of the typical energy gained from the electric field between
collisions, and the energy lost due to inelastic scattering. Since, in our 
model $\tau_{sc}$ is largely independent
of the electric field (it may acquire some corrections at large electric fields), 
the corrections to scaling due to
noise ($(T_{eff}\tau_{sc})^{2/3} \ll 1$) is subdominant in the quantum critical regime, with the dominant
scaling behavior being that derived in~\cite{Phillips04}. In the
quantum disordered regime however, direct coupling and noise effects
are found to be equally important. Current drift on the other hand
gives a correction which is smaller by an additional factor of
${\cal O}(1/E_F \tau_{sc})$  where $E_F$ is the fermi energy.

The paper is organized as follows. The model is presented in Section~\ref{model} and is treated within a
Keldysh path
integral approach which will allow us to study out-of-equilibrium effects. We first study the
equilibrium properties of the system by performing a
mean-field treatment in section~\ref{mfeq} which reveals a dissipation induced quantum critical point,
which can also be understood as a proximity effect.
A derivation of the
effective action for the superconducting fluctuations about the equilibrium ordered
side is presented in Appendix~\ref{flucord}, and the origin of a local
Caldiera-Leggett dissipation arising due to nonconserved particle number is highlighted.
Fluctuation about the nonequilibrium disordered state is studied in
section~\ref{neqfluc} and the new terms in the bosonic theory corresponding to
current noise and drift are derived.
Scaling equations for the gap and the current
are derived in section~\ref{scaling}. Many of the details of the derivation have been relegated
to the appendices. Finally we conclude in section~\ref{concl} where we discuss our results in 
the context of existing experiments.

\section{Model} \label{model}

\begin{figure}
\includegraphics[totalheight=3cm,width=6cm]{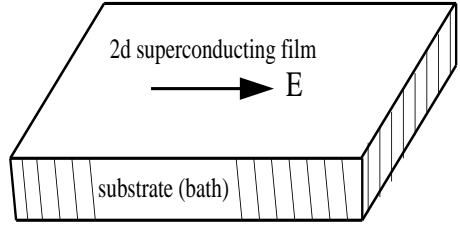}
\caption{A 2d itinerant electron system near a superconducting
instability and driven out of equilibrium by application of an in-plane
electric field. A steady state is reached via coupling to a normal metal substrate.
}
\label{schem}
\end{figure}

We consider a model of electrons in a $2$d layer that interact via a
short ranged attractive interaction responsible for a
superconducting instability and are coupled via tunneling to a
reservoir of non-interacting electrons. The Hamiltonian for the
system is
\begin{equation}
H = H_{bath} + H_{layer} + H_{layer-bath} \label{hm}\\
\end{equation}
where $H_{layer}$
is the interacting electron layer whose critical properties we
are interested in, $H_{bath}$ represents the reservoir, while $H_{layer-bath}$ represents the coupling
between the two.
\begin{eqnarray}
H_{layer} &=& \sum_{\sigma}\psi^{\dagger}_{\sigma}
\frac{1}{2m}\left(\frac{\vec{\nabla}}{i}- \frac{e}{\hbar c}\vec{A}\right)^2
\psi^{\dagger}_{\sigma} -
\lambda \psi^{\dagger}_{\uparrow} \psi^{\dagger}_{\downarrow} \psi_{\downarrow}\psi_{\uparrow}  \label{hsys}\\
H_{bath} &=& \sum_{k_z,k,\sigma} \epsilon^b_{k_z,k,\sigma} c_{k_z,k,\sigma}^{\dagger}
c_{k_z,k,\sigma} \label{hbath}\\
H_{layer-bath} &=& \sum_{\sigma,k_z,k}
\left(t
c^{\dagger}_{k_z,k,\sigma}
\psi_{k\sigma} + h.c.\right) \label{hcoup}
\end{eqnarray}
$\sigma$ is the spin label, $c$ represent the reservoir electrons, $k_z$ is the momentum
transverse to the superconductor-bath interface and is not conserved on tunneling, while $k$
is the momentum within the layer.
We assume the superconductor-bath interface
to be smooth, so that the in-plane momentum is conserved on tunneling. The schematic of
the model is shown in Fig~\ref{schem}.
In addition the electrons in the interacting layer are subjected to a
dc electric field which we represent via a vector potential $A= -c E t$. (We will set
$\hbar=1)$.

We write the Keldysh action for this model~\cite{Keldysh63,Kamenev04,Kamenev07},
\begin{equation}
Z_K = \int { {\cal D}}\left[ \psi_{\pm \sigma}, \bar{\psi}_{\pm \sigma},
c_{\pm,\sigma}, \bar{c}_{\pm,\sigma}\right]
e^{i \int_{-\infty}^{\infty} dt  d^dx \left(\left[L^e_- + L^{res}_{-}\right] -
\left[L^e_+ + L^{res}_{+}\right]\right)}
\end{equation}
where $\pm$ labels the Keldysh time-ordering, $L^{res}$ is the action for the reservoir electrons, while
$L^e$ is the Keldysh action for the layer electrons and the coupling with the reservoir.
\begin{eqnarray}
&&L^e_{\pm} = \sum_{\sigma}\bar{\psi}_{\pm \sigma} \left[i\frac{\partial}{\partial t} - \frac{1}{2m}
\left( \frac{\vec{\nabla}}{i} - \frac{e}{c}\vec{A}\right)^2 + \mu\right ] \psi_{\pm \sigma}
+ \lambda \bar{\psi}_{\pm \uparrow} \bar{\psi}_{\pm \downarrow} \psi_{\pm \downarrow} \psi_{\pm \uparrow}
- \sum_{\sigma} t \left[ \psi^{\dagger}_{\pm \sigma}(x) c_{\pm \sigma}(x) + h.c.\right]
\label{Le} \\
&&L^{res}_{\pm} = \sum_{\sigma} \bar{c}_{\pm \sigma}
\left[i\frac{\partial}{\partial t} - H_0 + \mu \right] c_{\pm \sigma} \label{Lres}
\end{eqnarray}
where $\mu$ is a chemical potential.

We perform a Hubbard Stratonovich decoupling of the attractive interaction
\begin{eqnarray}
\exp{\left(i \lambda \int dt d^dx \left[\bar{\Psi}_{- \uparrow} \bar{\Psi}_{- \downarrow} \Psi_{- \downarrow}
\Psi_{- \uparrow} - \bar{\Psi}_{+ \uparrow} \bar{\Psi}_{+ \downarrow} \Psi_{+ \downarrow}
\Psi_{+ \uparrow}  \right]\right)} = \int {{\cal D}}\left[\Delta_{\pm}, \Delta^*_{\pm} \right]
\exp{\left(-i\int dt d^d x \frac{|\Delta_-|^2 - |\Delta_+|^2}{\lambda}\right)} \nonumber \\
\exp{\left(i \int dt d^d x
\left( \Delta_{-} \bar{\Psi}_{-\uparrow} \bar{\Psi}_{-\downarrow}
+ \Delta_{-}^* \Psi_{-\downarrow} \Psi_{-\uparrow}  -
 \Delta_{+} \bar{\Psi}_{+\uparrow} \bar{\Psi}_{+\downarrow}
- \Delta_{+}^* \Psi_{+\downarrow} \Psi_{+\uparrow}\right)\right)}
\end{eqnarray}
and  in the process introduce the bosonic fields $\Delta_{\pm}$ which represent superconducting fluctuations.
Using Nambu notation $\Psi_{\pm} = \begin{pmatrix} \psi_{\pm \uparrow} \\ \bar{\psi}_{\pm \downarrow}\end{pmatrix},
\bar{\Psi}_{\pm} = \begin{pmatrix} \bar{\psi}_{\pm \uparrow} &\psi_{\pm \downarrow}\end{pmatrix},
\hat{c}_{\pm} = \begin{pmatrix} c_{\pm \uparrow} \\ \bar{c}_{\pm \downarrow}
\end{pmatrix},
\hat{\bar{c}}_{\pm} = \begin{pmatrix} \bar{c}_{\pm \uparrow} & c_{\pm \downarrow}
\end{pmatrix}
$
the Lagrangian becomes,
\begin{eqnarray}
&&L^e_{\pm} =  \bar{\Psi}_{\pm} \begin{pmatrix} i\frac{\partial}{\partial t} - \frac{1}{2m}
\left( \frac{\vec{\nabla}}{i} - \frac{e}{c}\vec{A}\right)^2 + \mu& \Delta_{\pm}
\\ \Delta^*_{\pm} & i\frac{\partial}{\partial t} + \frac{1}{2m}
\left( \frac{\vec{\nabla}}{i} + \frac{e}{c}\vec{A}\right)^2 -\mu
\end{pmatrix} \Psi_{\pm} \nonumber \\
&&-  t \left[\bar{\Psi}_{\pm}\begin{pmatrix} 1 &0 \\0& -1
\end{pmatrix}\hat{c}_{\pm} + h.c. \right]
- \frac{|\Delta_{\pm}|^2}{\lambda}
\label{Lenam}
\end{eqnarray}

The electronic degrees of freedom may now be formally integrated out, resulting in a Keldysh action entirely
in terms of the fluctuating fields $\Delta_{\pm}$. A rotation to retarded, advanced,
Keldysh space leads to~\cite{Kamenev04}
\begin{equation}
Z_K = \int {{\cal D}}\left[ \Delta_{q,cl},\Delta^*_{q,cl}\right] \exp{\left(Tr \ln {\cal G}^{-1} \right)}
\exp{\left(-i \int dt d^d x
\left[\frac{2 \Delta_q^* \Delta_{cl} + 2 \Delta_{cl}^* \Delta_q}{\lambda}\right]\right)}
\label{ZK3}
\end{equation}
where $\Delta_{q}= \frac{\Delta_{-}-\Delta_{+}}{2}$,
$\Delta_{cl} = \frac{\Delta_- + \Delta_+}{2}$ are respectively the quantum and classical components of the
fluctuating fields.
${\cal G}$ is a $ 4 \times 4$ matrix in Nambu and Keldysh ($\tau_{x,y,z}$) space which
obeys the Dyson equation
\begin{eqnarray}
&{\cal{G}}^{-1} = {\cal{G}}_0^{-1} + \begin{pmatrix}0 & \Delta_{cl} \\ \Delta_{cl}^* & 0 \end{pmatrix}
\otimes \tau_0 +
\begin{pmatrix}0 & \Delta_{q} \\ \Delta_{q}^* & 0  \end{pmatrix} \otimes \tau_x \label{Gexp}
\end{eqnarray}
where ${\cal{G}}_0$ is the exact Green's function for non-interacting electrons coupled to
reservoirs and subjected to an external electric field. The full Green's function
${\cal{G}}$ may be written as follows in Nambu and Keldysh space,
${\cal G} = \begin{pmatrix}{\cal G}^R& {\cal G}^K\\0&{\cal G}^A \end{pmatrix}$
where the ${\cal G}^{R,A,K}$ are the following $2 \times 2$ matrices,
\begin{equation}
\frac{1}{2}{\cal{G}}^R(t,t^{\prime}) = -i \langle \Psi_{cl}(t)\bar{\Psi}_q(t^{\prime})\rangle =
\frac{1}{2}\begin{pmatrix} G^R &  F^R\\ \bar{F}^R & \bar{G}^R  \end{pmatrix} \nonumber 
\end{equation}
\begin{eqnarray}
\frac{1}{2}{\cal{G}}^K(t,t^{\prime}) = -i \langle \Psi_{cl}(t)\bar{\Psi}_{cl}(t^{\prime})\rangle =
\frac{1}{2}\begin{pmatrix} G^K &  F^K\\ \bar{F}^K & \bar{G}^K  \end{pmatrix} \nonumber 
\end{eqnarray}
and the retarded Green's functions are defined as,
\begin{eqnarray}
&&G^R(x,t;x^{\prime},t^{\prime}) = -i \theta(t-t^{\prime})\langle\{\psi_{\uparrow}(x,t),
\bar{\psi}_{\uparrow}(x^{\prime},t^{\prime})\}\rangle = G^R_{\uparrow}(x,t;x^{\prime},t^{\prime}) \\
&& \bar{G}^R(x,t;x^{\prime},t^{\prime}) = -i \theta(t-t^{\prime})\langle\{\bar{\psi}_{\downarrow}(x,t),
{\psi}_{\downarrow}(x^{\prime},t^{\prime})\}\rangle = - G^A_{\downarrow}(x^{\prime}t^{\prime};x,t)\\
&&F^R(x,t;x^{\prime},t^{\prime}) = -i \theta(t-t^{\prime})\langle\{\psi_{\uparrow}(x,t),
{\psi}_{\downarrow}(x^{\prime},t^{\prime})\}\rangle \\
&& \bar{F}^R(x,t;x^{\prime},t^{\prime}) = -i \theta(t-t^{\prime})\langle\{\bar{\psi}_{\downarrow}(x,t),
\bar{\psi}_{\uparrow}(x^{\prime},t^{\prime})\}\rangle
\end{eqnarray}
and the Keldysh Green's functions are
\begin{eqnarray}
&&G^K(x,t;x^{\prime},t^{\prime}) = -i \langle\left[\psi_{\uparrow}(x,t),
\bar{\psi}_{\uparrow}(x^{\prime},t^{\prime})\right]\rangle = G^K_{\uparrow}(x,t;x^{\prime},t^{\prime}) \\
&& \bar{G}^K(x,t;x^{\prime},t^{\prime}) = -i\langle\left[\bar{\psi}_{\downarrow}(x,t),
{\psi}_{\downarrow}(x^{\prime},t^{\prime})\right]\rangle = - G^K_{\downarrow}(x^{\prime}t^{\prime};x,t)\\
&&F^K(x,t;x^{\prime},t^{\prime}) = -i \langle\left[\psi_{\uparrow}(x,t),
{\psi}_{\downarrow}(x^{\prime},t^{\prime})\right]\rangle \\
&& \bar{F}^K(x,t;x^{\prime},t^{\prime}) = -i \langle\left[\bar{\psi}_{\downarrow}(x,t),
\bar{\psi}_{\uparrow}(x^{\prime},t^{\prime})\right]\rangle
\end{eqnarray}

\section{Mean field treatment in equilibrium ($E= 0$)} \label{mfeq}

The mean-field equations may be obtained by minimizing Eq.~\ref{ZK3} with respect to
the quantum ($\Delta_{q}$) and classical ($\Delta_{cl}$) fluctuations of the
order-parameter.
A Ginzburg-Landau action is then obtained by expanding the Keldysh functional in
fluctuations about the mean field solution.
We first outline these steps for the equilibrium case {\sl i.e.}, $ A = E t = 0$, before turning to
the nonequilibrium case.

In equilibrium the single particle Green's function ${\cal G}_0$ may be easily obtained. In Fourier space
the retarded Green's functions are,
\begin{eqnarray}
{G}_{0R}^{-1} = \omega - \xi_k - \Sigma^R\\
\bar{{G}}_{0R}^{-1} = \omega + \xi_k - \bar{\Sigma}^R
\end{eqnarray}
where $\xi_k = \epsilon_k - \mu$, and the self-energies $\Sigma$ arise due to coupling to reservoirs and have the form,
\begin{eqnarray}
\Sigma^R = \sum_{k_z}\frac{t^2}{\omega - \epsilon^b_{k_z} - \epsilon^b_{k} + \mu +i \delta} \simeq -i\Gamma\\
\bar{\Sigma}^R = \sum_{k_z}\frac{t^2}{\omega + \epsilon^b_{k_z} + \epsilon^b_{k} -\mu + i \delta} \simeq -i\Gamma
\end{eqnarray}
where $\Gamma = \pi \rho t^2$ , $\rho$ being the density of states of the reservoirs. We have
taken the reservoir dispersion in Eq.~\ref{hbath} to be $\epsilon^b_{k_z,k}= \epsilon^b_{k_z} + \epsilon^b_k$.
Note that we will interchangeably use the notation
\begin{equation}
\tau_{sc} = \frac{1}{2\Gamma}
\end{equation}
to represent the typical escape time into the reservoirs.

For a reservoir in equilibrium at temperature $T$, the Keldysh self energies of the layer
electrons due to coupling to the reservoir obey the fluctuation-dissipation theorem,
\begin{eqnarray}
\Sigma^K = -2i\Gamma \tanh \frac{\omega}{2T} \\
\bar{\Sigma}^K = -2i \Gamma \tanh \frac{\omega}{2T}
\end{eqnarray}
Moreover,
\begin{eqnarray}
G^K_0 = G^R_0 \Sigma^K G^A_0 \\
\bar{G}^K_0 = \bar{G}^R_0 \bar{\Sigma}^K \bar{G}^A_0
\end{eqnarray}
It therefore follows that
\begin{equation}
{\cal G}_{R0}^{-1} =
\begin{pmatrix}\omega - \xi_{k} + i\Gamma & 0
\\ 0 & \omega + \xi_{k} + i \Gamma \end{pmatrix} => \\
{\cal G}_{R0} = \frac{1}{\left( \omega + i \Gamma\right)^2
-\xi_k^2}\begin{pmatrix}\omega + \xi_k + i\Gamma & 0
\\ 0 & \omega - \xi_k + i \Gamma \end{pmatrix}
\end{equation}
and the fluctuation-dissipation theorem is obeyed so that,
\begin{equation}
{\cal{G}}_{0K} =
-2i \Gamma \left[\tanh \frac{\omega}{2T} \right]{\cal G}_{0R} {\cal G}_{0A} =
\left({\cal G}_{0R} - {\cal G}_{0A} \right)
\tanh \frac{\omega}{2T}
\end{equation}
The single particle Green's functions computed above for electrons coupled to reservoirs
seem identical to those
for electrons scattering elastically off impurities. However, the difference between the two systems 
will be apparent in the single particle level in the next section when an electric field is 
applied. In that case, while there is no steady state for electrons scattering off static impurities,
the coupling to a reservoir in our model 
will be shown to provide an inelastic mechanism which will allow the system to reach 
a nonequilibrium steady state. The difference between the two systems is also apparent in equilibrium when 
electronic response and correlation functions are computed. For a disordered system appropriate disorder
averaging gives answers which are consistent with a closed system characterized by conserved particle number. 
For our system the response and correlation functions (computed in Section~\ref{neqfluc}) will reflect the 
fact that the system is open since electrons can escape into the reservoir.

We now expand the Trln in Eq~\ref{ZK3} about $\Delta_q \rightarrow
\bar{\Delta}_q + \Delta_q$, $\Delta_{cl} \rightarrow \Delta_{0} +
\Delta_{cl}$, and determine $\bar{\Delta}_q, \Delta_0$ so that the
resultant action is a minimum with respect to both quantum and
classical fluctuations of the order-parameter. By choosing
$\bar{\Delta}_q=0$, the resultant Keldysh action is automatically
minimized with respect to classical fluctuations of the
order-parameter~\cite{Kamenev04}, whereas $\Delta_0$ will be
determined by minimizing with respect to the quantum fluctuations.
Thus the mean field Green's function is the matrix
\begin{equation}
{\cal{G}}_{mf}^{-1} = {\cal{G}}_0^{-1} + \begin{pmatrix}0 & \Delta_{0} \\ \Delta_{0} & 0 \end{pmatrix}
\otimes \tau_0  \label{Gmfdef}
\end{equation}
where the retarded component is
\begin{eqnarray}
{\cal G}_{Rmf}^{-1} = \begin{pmatrix}\omega - \xi_k + i\Gamma & \Delta_0
\\ \Delta_0 & \omega + \xi_k + i \Gamma \end{pmatrix} => \\
{\cal G}_{Rmf} = \frac{1}{\left( \omega + i \Gamma\right)^2
- \Delta_0^2 -\xi_k^2}\begin{pmatrix}\omega + \xi_k + i\Gamma & -\Delta_0
\\ -\Delta_0 & \omega - \xi_k + i \Gamma \end{pmatrix} \label{Grmf}
\end{eqnarray}
and the Keldysh component is,
\begin{eqnarray}
&{\cal{G}}_{Kmf} = -2i \Gamma \left[\tanh \frac{\omega}{2T} \right]{\cal G}_{Rmf} {\cal G}_{Amf}
= \tanh \frac{\omega}{2T} \left[G_{Rmf} - G_{Amf}\right]\\
&=\left[\tanh \frac{\omega}{2T} \right]
\begin{pmatrix}\frac{\omega + \xi_k + i \Gamma}{\left(\omega + i \Gamma \right)^2 -\xi_k^2 - \Delta_0^2}
- \frac{\omega + \xi_k - i \Gamma}{\left(\omega-i\Gamma\right)^2-\xi_k^2 -\Delta_0^2 }
&-\Delta_0\left(\frac{1}{\left(\omega + i \Gamma \right)^2 -\xi_k^2 - \Delta_0^2}
- \frac{1}{\left(\omega-i\Gamma\right)^2-\xi_k^2 -\Delta_0^2} \right) \\
-\Delta_0\left(\frac{1}{\left(\omega + i \Gamma \right)^2 -\xi_k^2 - \Delta_0^2}
- \frac{1}{\left(\omega-i\Gamma\right)^2-\xi_k^2 -\Delta_0^2}\right)
& \frac{\omega - \xi_k + i \Gamma}{\left(\omega + i \Gamma \right)^2 -\xi_k^2 - \Delta_0^2}
- \frac{\omega - \xi_k - i \Gamma}{\left(\omega-  i\Gamma \right)^2-\xi_k^2 -\Delta_0^2 }
\end{pmatrix}\\
&= \begin{pmatrix} G^K & F^K \\ \bar{F}^K & \bar{G}^K \end{pmatrix}
\label{GKmf}
\end{eqnarray}

$\Delta_0$ is obtained by minimizing Eq.~\ref{ZK3} with respect to $\Delta_q^*$
which leads to the self-consistent gap equation,
\begin{equation}
\frac{-2i \Delta_{0}}{\lambda} + Tr\left[F_K\right] = 0 \label{selfcons}
\end{equation}
The above equation, together with Eq.~\ref{GKmf} implies the following condition for a superconducting
instability
\begin{equation}
\frac{-2i}{\lambda} - \frac{1}{L^d}\sum_k \int_{-\Omega_{bcs}}^{\Omega_{bcs}} \frac{d\omega}{2\pi}
\tanh\frac{\omega}{2T} \left[\frac{1}{\left(\omega + i \Gamma
\right)^2 -\xi_k^2} -
\frac{1}{\left(\omega-i\Gamma\right)^2-\xi_k^2} \right] = 0
\label{Gameq}
\end{equation}
For $\Gamma=0, T\neq 0$, the above is the usual equation for the BCS mean-field transition temperature.
On the other hand, at $T=0, \Gamma \neq 0$,
Eq.~\ref{Gameq} yields a critical value of the dissipation $\Gamma_{c}$, above which
the system becomes normal
\begin{equation}
\frac{1}{\nu \lambda} = \ln\frac{\Omega_{bcs}}{\Gamma_c}
\end{equation}
where $\nu$ is the single particle density of states at the Fermi energy.
The destruction of superconductivity due to large $\Gamma$
can be understood in a simple way as a proximity effect.
$\Gamma$ measures the amount by which the states in the 2d layer broaden due to hybridization with the
normal metal leads. Thus the larger is $\Gamma$ the more the states in the layer acquire the property of
the underlying normal metal. For an early reference on similar proximity effect induced
destruction of superconductivity in a thin superconducting film
deposited on a normal metal see~\cite{DeGennes64}.

A second interesting feature of this model
is a gapless spectrum even when there are nonzero superconducting correlations.
To see this we evaluate the quasi-particle density of states per spin direction $N(\omega) = \frac{i}{2\pi}\sum_k
\left(G^R - G^A\right)(k,\omega)$, and find that at it is nonvanishing for $\omega < \Delta_0$. In particular
at zero frequency
\begin{eqnarray}
N(\omega=0) = \frac{\nu \Gamma}{\sqrt{\Gamma^2 + \Delta_0^2}}
\end{eqnarray}
This appearance of gapless  superconductivity
is also a typical property of superconductor-metal interfaces~\cite{Park}.

To fully understand the equilibrium properties a derivation of the superconducting action on
the ordered side is presented in Appendix~\ref{flucord}. To keep the notation simple, this
derivation has been done for the partition function, and the
effective action for the phase fluctuations in imaginary time is found to be
(Eq.~\ref{Sord1})
\begin{equation}
S = \int d\tau \int d^d x\left[c_1 \left(\partial_{\tau}\theta\right)^2 + c_2
\left(\vec{\nabla} \theta - \frac{e}{c}\vec{A}\right)^2\right] + S_{\Sigma}
\end{equation}
where
\begin{equation}
S_{\Sigma} = \frac{g}{2\pi}
\int d^d x \int d\tau \int d\tau^{\prime}
\left(\frac{\theta(x,\tau)-\theta(x,\tau^{\prime})}{\tau - \tau^{\prime}}\right)^2 \label{Sdiss2a}
\end{equation}
(Eq.~\ref{Sdiss2}).
The distinguishing feature is the dissipative term $S_{\Sigma}$
which arises because the
superconducting layer is characterized by non-conserved particle number. Since the physical quantity
is the voltage fluctuations $V = \partial_{\tau}\theta $ in the
superconducting layer relative to the normal substrate, it is instructive to
rewrite Eq.~\ref{Sdiss2a} after an integration by parts,
\begin{eqnarray}
S_{\Sigma} = -\frac{g}{\pi}
\int d^d x \int d\tau \int d\tau^{\prime} \partial_{\tau}\theta(x,\tau) \ln{|\tau-\tau^{\prime}|}
\partial_{\tau^{\prime}}\theta(x,\tau^{\prime})
\end{eqnarray}

\section{Mean field treatment out of equilibrium} \label{mfneq}

We now turn to a mean-field treatment of the out of equilibrium
current carrying case. A mean-field approach relies on the
assumption that even in the presence of current flow the transition
from the normal to the superconducting side is second order. In what
follows the mean-field phase boundary in the current vs. equilibrium
superconducting gap
plane will be derived coming in
from the disordered side. This will be followed by a discussion of
the validity of mean-field.

The self-consistent mean-field equations for the current carrying
case is still given by Eq.~\ref{Gmfdef}, however we now have to
evaluate ${\cal G}_0$ which is the Green's functions for the layer
electrons coupled to external reservoirs and subjected to an
electric field~\cite{Mitra08a}. An analytic solution may be obtained
in the limit where $1/E_F \tau_{sc} \ll 1$ and $\left(\frac{T_{eff}}{E_F}\right)T_{eff}\tau_{sc}\ll
1$, where $\tau_{sc} = 1/(2\Gamma)$ is the typical escape time into
the reservoir, and $T_{eff} = e E v_F \tau_{sc}$ is an effective
temperature that characterizes the steady state distribution
function of the layer electrons. The details are presented in
Appendix~\ref{steadystate}. Within this approximation the mean-field
green's functions ${\cal G}^{R,A}_{mf}$ remain unchanged and are
still given by Eq.~\ref{Grmf}, while ${\cal G}^K_{mf}$ changes due
to a nonequilibrium electronic distribution function. In what 
follows, both in the mean-field treatment of this section, and the study of
fluctuations in the next section and in Appendix~\ref{POL}, we will make the additional assumption that 
$T_{eff}\tau_{sc} \ll 1$.

Using the expression for the
distribution function in Eq.~\ref{fs1},~\ref{fssolph},~\ref{fasolph},
the self-consistent gap equation Eq~\ref{selfcons} becomes,
\begin{eqnarray}
\frac{2i}{\nu \lambda} = -\int_{-\pi}^{\pi}\frac{d\theta}{2\pi} \int_0^{\Omega_{bcs}} \frac{d\omega}{\pi}
\left(1 - e^{\frac{-\omega}{T_{eff}|\cos\theta|}} \right) \int_{-\infty}^{\infty} d \xi
\left[\frac{1}{(\omega + i \Gamma)^2 -\xi^2 -\Delta_0^2}
- \frac{1}{(\omega - i \Gamma)^2 -\xi^2 -\Delta_0^2}\right]
\label{selfconeq}
\end{eqnarray}
We may approximate $\int_{-\pi}^{\pi}\frac{d\theta}{2\pi}
\int_0^{\Omega_{bcs}} \frac{d\omega}{\pi} \left(1 -
e^{\frac{-\omega}{T_{eff}|\cos\theta|}} \right) \simeq
\int_{T_{eff}}^{\Omega_{bcs}} \frac{d\omega}{\pi}$. Defining
$T_{c,eff}$ as the critical value of the current induced effective
temperature for which $\Delta_0=0$ in Eq.~\ref{selfconeq}, and
relating $\frac{1}{\nu \lambda}$ to the gap $\Delta_{eq}$ in
equilibrium we find
\begin{eqnarray}
\frac{2}{\pi} \int_0^{\Omega_{bcs}} \frac{d \xi}{\sqrt{\xi^2 +
\Delta_{eq}^2}} \arctan{\frac{\sqrt{\xi^2+\Delta_{eq}^2}}{\Gamma}}=
\ln \frac{\Omega_{bcs}}{\sqrt{T_{c,eff}^2 + \Gamma^2}}
\label{Tcneqdef}
\end{eqnarray}
An approximate solution to the above equation is $T_{c,eff} \simeq
\Delta_{eq}/\sqrt{2}$. Note that this current induced loss of order
is a heating effect arising due to a highly broadened electron
distribution function, and is not the same as the Landau-criterion
for the critical current for breakdown of superfluidity.

We now turn to the discussion of the validity of the above
mean-field treatment. Firstly, within mean-field $T_{eff}$ arises
solely out of noise due to normal electron current, and leaves out
the fact that current due to superconducting fluctuations also
contribute to noise that can modify $T_{eff}$. Secondly, other
scenarios for a current induced transition are possible. For
example, on the superconducting side there is no dissipative
current, so that $T_{eff}=0$. In this case it is possible to have a
supercurrent induced first order transition from a superconducting
state to a normal or resistive state as discussed
in~\cite{Kopnin84,Polkovnikov05}. Whether the actual transition is a
second order heating effect as predicted by the mean-field treatment
on the disordered side or a first order transition depends on
whether the critical current for the first order transition is
larger or smaller than the current corresponding to $T_{c,eff}$. In
general this is a complex question that we do not address further.

Instead, in what follows we will derive a nonequilibrium
Ginzburg-Landau theory for the superconducting fluctuations on the
normal side where most of the current is due to normal electrons so
that there is a well defined $T_{eff}$. We will then use this to
study how the gap and the current due to superconducting
fluctuations scale due to electric field close to the quantum
critical point and outside the fluctuation dominated Ginzburg
regime.

\section{Fluctuation about nonequilibrium disordered state:
Derivation of the Keldysh Ginzburg-Landau functional} \label{neqfluc}

We now turn to the discussion of fluctuations about the mean-field disordered state
($\Delta_0=0$ in Eq.~\ref{Gmfdef}).
In doing so we will also highlight the difference between how the
electric field affects magnetic-fluctuations~\cite{Mitra08a} and superconducting fluctuations.

Expanding the $Tr\ln$ in Eq.~\ref{ZK3} in the usual way
$Tr\ln {\cal G}^{-1} = Tr\ln {\cal G}_0^{-1} + Tr{\cal G}_0 {\hat \Delta} -\frac{1}{2}
Tr{\cal G}_0 {\hat \Delta} {\cal G}_0 {\hat \Delta} + \ldots $
where
\begin{equation}
{\hat \Delta} = \begin{pmatrix}0 & \Delta_{cl} \\ \Delta_{cl}^* & 0 \end{pmatrix}
\otimes \tau_0 +
\begin{pmatrix}0 & \Delta_{q} \\ \Delta_{q}^* & 0  \end{pmatrix} \otimes \tau_x
\end{equation}
one obtains an effective action
\begin{equation}
Z_K = \int {\cal D}\left[\Delta_{q,cl},\Delta_{q,cl}^*\right] e^{-i S_K^2 - i S_K^3 - i S_K^4 + \ldots}
\label{ZK3eq}
\end{equation}
with
\begin{equation}
S_K^2 = Tr
\begin{pmatrix} \Delta_q^* & \Delta_{cl}^*\end{pmatrix}
\begin{pmatrix}\Pi^K_{G\bar{G}} & \frac{2}{\lambda} + \Pi^R_{G\bar{G}}
\\\frac{2}{\lambda} + \Pi^A_{G\bar{G}}  & 0 \end{pmatrix}  \begin{pmatrix} \Delta_q \\
\Delta_{cl}\end{pmatrix}
\end{equation}
In position and time space $1 = x, t $ and $2 = x^{\prime} t^{\prime} $
\begin{eqnarray}
&&\Pi^R_{G\bar{G}}(1,2) = -i\left[G_{0R}(1,2)
\bar{G}_{0K}(2,1) + G_{0K}(1,2) \bar{G}_{0A}(2,1) \right] \\
&&\Pi^K_{G\bar{G}}(1,2) = -i\left[G_{0K}(1,2)
\bar{G}_{0K}(2,1) + G_{0R}(1,2) \bar{G}_{0A}(2,1) +
G_{0A}(1,2) \bar{G}_{0R}(2,1)  \right]
\end{eqnarray}

Note that on the disordered side, terms cubic order in the superconducting fluctuations are absent($S_K^3 = 0$),
while $S_K^4$ has the form~\cite{Mitra06}
\begin{eqnarray}
S_K^4 = \sum_{i=1\ldots 4} u_i \Delta^{*i}_q \Delta_{cl}^{4-i} + c.c.
\label{SK4}
\end{eqnarray}
We will treat $S_K^4$ only within a one-loop mean-field approximation. For this only the coupling constant
$u_1$ will play a role.

Since $ \bar{G}_R(1,2) = -G_A(2,1) \Rightarrow {\bar{G}}_R(\vec{k},\omega)
= - {G}_A(-\vec{k},-\omega), \bar{G}_K(1,2) = -G_K(2,1) \Rightarrow
{\bar{G}}_K(\vec{k},\omega) = - {G}_K(-\vec{k},-\omega) $, we may write
\begin{eqnarray}
Tr[\Delta_q^* \Pi^R_{G\bar{G}} \Delta_{cl}]
= iTr\Delta_q^*(1) \left[G_{0R}(1,2)
{G}_{0K}(1,2) + G_{0K}(1,2) {G}_{0R}(1,2) \right] \Delta_{cl}(2)\\
Tr[\Delta_q^* \Pi^K_{G\bar{G}} \Delta_{q}]
= iTr \Delta_q^*(1) \left[G_{0K}(1,2)
{G}_{0K}(1,2) + G_{0R}(1,2) {G}_{0R}(1,2) +
G_{0A}(1,2) {G}_{0A}(1,2)\right]\Delta_{q}(2)
\end{eqnarray}

The coefficients $\Pi$ depend explicitly on the electric field. Since we use the Gauge
$\vec{A} = - c \vec{E} t$,
it is most convenient to go into momentum and time space,
\begin{eqnarray}
Tr[\Delta_q^*\Pi^R_{G\bar{G}} \Delta_{cl}]
= iTr\Delta_q^*(-\vec{q},t_1) \left[G_{0R}(\vec{p}+\vec{q};t_1,t_2)
{G}_{0K}(-\vec{p};t_1,t_2) + G_{0K}(\vec{p}+\vec{q};t_1,t_2) {G}_{0R}(-\vec{p};t_1,t_2) \right] 
\Delta_{cl}(\vec{q},t_2)
\label{exps1}
\end{eqnarray}
Note that for magnetic fluctuations, the above expression would have had the form~\cite{Mitra08a}
\begin{eqnarray}
Tr[m_q\Pi^R_{G\bar{G}}m_{cl}]
= iTrm_q^*(-\vec{q},t_1) \left[G_{0R}(\vec{p}+\vec{q};t_1,t_2)
{G}_{0K}(\vec{p};t_2,t_1) + G_{0K}(\vec{p}+\vec{q};t_1,t_2) {G}_{0A}(\vec{p};t_2,t_1) \right] 
m_{cl}(\vec{q},t_2)
\label{expm1}
\end{eqnarray}
As shown in Appendix~\ref{steadystate}, if the single particle Green's functions are written in terms of the
canonical momentum $\vec{k} = \vec{p} + e \vec{E} T$ 
where $T = \frac{t_1+t_2}{2}$, they become time translationally
invariant.
Thus for superconducting fluctuations one may write Eq.~\ref{exps1} as
\begin{eqnarray}
Tr[\Delta_q^*\Pi^R_{G\bar{G}} \Delta_{cl}]
= iTr\Delta_q^*(-\vec{q},t_1) \{G_{0R}(\vec{p}+\vec{q} + e \vec{E} T;t_1-t_2)
{G}_{0K}(-\vec{p} + e \vec{E} T; t_1-t_2) \nonumber \\
+ G_{0K}(\vec{p}+\vec{q} + e \vec{E} T;t_1-t_2) 
{G}_{0R}(-\vec{p}+ e \vec{E} T;t_1-t_2) \} \Delta_{cl}(q,t_2)
\label{exps2}
\end{eqnarray}
In terms of the canonical momentum $k = p + e E T$, the above becomes
\begin{eqnarray}
Tr[\Delta_q^*\Pi^R_{G\bar{G}} \Delta_{cl}]
= iTr\Delta_q^*(-\vec{q},t_1) \{G_{0R}(\vec{k}+\vec{q} ;t_1-t_2)
{G}_{0K}(-\vec{k} + 2 e \vec{E} T; t_1-t_2) \nonumber \\
+ G_{0K}(\vec{k} +\vec{q};t_1-t_2) {G}_{0R}(-\vec{k} + 2 e \vec{E} T;t_1-t_2) \} \Delta_{cl}(q,t_2)
\label{exps3}
\end{eqnarray}
or shifting variables $\vec{k} \rightarrow \vec{k} + 2 e \vec{E} T$ one may write
\begin{eqnarray}
Tr[\Delta_q^*\Pi^R_{G\bar{G}} \Delta_{cl}]
= iTr\Delta_q^*(-\vec{q},t_1) \{G_{0R}(\vec{k}+\vec{q} + 2e \vec{E} T;t_1-t_2)
{G}_{0K}(-\vec{k}; t_1-t_2) \nonumber \\
+ G_{0K}(\vec{k}+\vec{q} + 2 e \vec{E} T;t_1-t_2) {G}_{0R}(-\vec{k};t_1-t_2) \} \Delta_{cl}(q,t_2)
\label{exps4}
\end{eqnarray}

Following the same steps for magnetic fluctuations we get
\begin{eqnarray}
&&Tr[m_q\Pi^R_{G\bar{G}}m_{cl}]
= iTrm_q^*(-\vec{q},t_1) \label{expm2}\\
&&\left[G_{0R}(\vec{p}+\vec{q} + e \vec{E} T;t_1-t_2)
{G}_{0K}(\vec{p} + e \vec{E} T,t_2-t_1) 
+ G_{0K}(\vec{p}+\vec{q} + e \vec{E} T;t_1-t_2) {G}_{0A}(\vec{p} + e \vec{E} T;t_2-t_1) \right] 
m_{cl}(\vec{q},t_2)
\label{expm3}
\nonumber
\end{eqnarray}
Rewriting the above in terms of the canonical momentum $\vec{k} = \vec{p} + e \vec{E} T$, one finds,
\begin{eqnarray}
&&Tr[m_q\Pi^R_{G\bar{G}}m_{cl}]
= iTrm_q^*(-\vec{q},t_1) \label{expm4}\\
&&\left[G_{0R}(\vec{k}+\vec{q};t_1-t_2)
{G}_{0K}(\vec{k},t_2-t_1) + G_{0K}(\vec{k}+\vec{q};t_1-t_2) 
{G}_{0A}(\vec{k};t_2-t_1) \right] m_{cl}(\vec{q},t_2)
\nonumber
\end{eqnarray}

Thus Eq.~\ref{expm4} and~\ref{exps4} highlight the difference
between the coupling of the electric field to the magnetic and
superconducting order parameters. In Eq.~\ref{expm4}, all dependence
of the electric field is via the modification of the Green's
functions $G^{R,K}$ at steady state, and there is no direct coupling
between the electric field and the order-parameter. On the other
hand Eq.~\ref{exps4} depends on the combination $(\vec{q} + 2 e \vec{E} T)$
which is the usual minimal coupling of the charged superconducting
fluctuation and an external electric field.

Thus to summarize, upto quadratic order, the Keldysh action for
superconducting fluctuations in the presence of an electric field
may be written as
\begin{eqnarray}
&&S_K^2 = \int dt_1 \int dt_2 \sum_{\vec{q}} 
\label{SK2E}\\
&&\begin{pmatrix} \Delta_q^*(-\vec{q},t_1) & \Delta_{cl}^*(-\vec{q},t_1)\end{pmatrix}
\begin{pmatrix}\Pi^K_{G\bar{G}}\left(\vec{q} + 2 e \vec{E} T, t_1-t_2 \right) & \frac{2}{\lambda}
\delta(t_1-t_2) + \Pi^R_{G\bar{G}}
\left(\vec{q} + 2 e \vec{E} T, t_1-t_2 \right)
\\
\frac{2}{\lambda}\delta(t_1-t_2) + \Pi^A_{G\bar{G}}\left(\vec{q} + 2 e \vec{E} T, t_1-t_2 \right)
& 0 \end{pmatrix}  \begin{pmatrix} \Delta_q(\vec{q},t_2) \\
\Delta_{cl}(\vec{q},t_2)\end{pmatrix}
\nonumber
\end{eqnarray}
where $T = \frac{t_1+t_2}{2}$. We now discuss the coefficients
$\Pi^{R,A,K}$ and highlight the appearance of new current dependent
terms that were missed in previous phenomenological treatments.

We expand the $\Pi$ bubbles in powers of $\left(\vec{q} + 2 e \vec{E} T \right)$ to obtain,
\begin{eqnarray}
&&\Pi^R\left(\vec{q} + 2 e \vec{E} T; t_1-t_2 \right) = \left[\Pi^0_R(t_1-t_2)
 +\vec{E}\cdot\left(\vec{q}+2e\vec{E}T\right) \Pi_1^R(t_1-t_2)
+\left(\vec{q} + 2 e \vec{E} T \right)^2 \Pi_2^R(t_1-t_2)+ \ldots \right] \label{PiRexp}\\
&&\Pi^K\left(\vec{q} + 2 e \vec{E} T; t_1-t_2 \right) = \Pi^K_0(t_1-t_2) + {\cal
O}((\vec{q} + 2 e \vec{E} T)^2) \label{PiKexp}
\end{eqnarray}
It is convenient to Fourier transform the above expressions so that
\begin{eqnarray}
&&\Pi^R\left(\vec{q} + 2 e \vec{E} T; t_1-t_2 \right)
= \int \frac{d\Omega}{2\pi} e^{-i \Omega (t_1-t_2)}\left[\tilde{\Pi}^R_0(\Omega)
+ \vec{E}\cdot\left(\vec{q}+2e\vec{E}T\right) \tilde{\Pi}_1^R(\Omega)  
+\left(\vec{q} + 2 e \vec{E} T \right)^2 \tilde{\Pi}^R_2(\Omega) \right]
\end{eqnarray}
Each of $\tilde{\Pi}^R_{0,1}(\Omega)$ can be evaluated as a power
series in $\Omega$ (see Appendix~\ref{POL} for details). Keeping
terms to ${\cal O}(\Omega, \left( \vec{q} + 2 e \vec{E} T\right)^2)
$ one obtains,
\begin{eqnarray}
\delta(t_1 -t_2) +
\frac{\lambda}{2}
\Pi^R\left(\vec{q} + 2 e \vec{E} T; t_1-t_2 \right)  = \delta(t_1 -t_2)\left[
\eta \left(\frac{\partial}{\partial t_1} -i \tau_{sc}\frac{e\vec{E}
\cdot\left(\vec{q}+2e\vec{E}T\right)}{m}\right)+
\delta + \gamma \left( \vec{q} + 2 e \vec{E} T\right)^2 + \ldots \right] \label{PiRexp2}
\end{eqnarray}
where, as derived in Appendix~\ref{POL}
\begin{eqnarray}
&&\eta = \nu \lambda \tau_{sc} \label{alpha} \\
&&\delta = 1 + \frac{\lambda}{2} {\sl Re} \left[ \tilde{\Pi}^R(0,0) \right] \label{gapdef}  \\
&&\gamma = \lambda \nu \frac{\mu}{4 m \Gamma^2}\label{gamma}
\end{eqnarray}
The first term on the r.h.s of Eq.~\ref{PiRexp2} is the overdamped dynamics associated with non-conservation of
particle number, while the second term is of the form $\vec{v}_D\cdot\left(\vec{q} + 2e \vec{A} t\right)$
and represents current induced drift at velocity
\begin{equation}
v_D = \frac{\tau_{sc}e E}{m} \label{vDdef}
\end{equation}
The difference with~\cite{Phillips04} is the appearance of the above
drift term, along with a change in the noise properties of the
reservoir (represented by $\Pi^K$) due to current flow. In
particular, we find the following electric-field dependence of
$\Pi^K$ in $2d$ (see Appendix~\ref{POL} for details)
\begin{equation}
\Pi^K(\Omega) = -4i\nu\tau_{sc}\left[|\Omega| + T_{eff} \int_{-\pi}^{\pi}
\frac{d\phi}{2\pi} |\cos\phi| e^{-\frac{|\Omega|}{T_{eff}|\cos\phi|}}
\right] \label{PiKneq}
\end{equation}
as opposed to a current independent $\Pi^K(\Omega) \propto
|\Omega|$, in the model studied in~\cite{Phillips04}.

Note that for a 1d system, the structure of $\Pi^{R,A}$ remain the
same as in 2d, while $\Pi^K$ acquires the form in Eq.~\ref{PiK1d}.
Qualitatively it has the same structure as Eq.~\ref{PiKneq} in that
$\Pi^K_{1d} \propto |\Omega|$ when $|\Omega| > T_{eff}$, and
$\Pi^K_{1d} \propto T_{eff}$ when $\Omega = 0$.

We now turn to the evaluation of the gap-equation and the current due to
superconducting fluctuations to see what role these new terms due to
current induced noise and drift play.

\section{Evaluation of self-consistent gap and current due to superconducting fluctuations} \label{scaling}

In order to derive the self-consistent gap equation and the
fluctuation conductivity, as in~\cite{Phillips04} we will work to
quadratic order (Eq.~\ref{SK2E}), treating the quartic term in
superconducting fluctuations (Eq.~\ref{SK4}) within a one-loop
mean-field approximation.

We may define the retarded, advanced and Keldysh component of the Green's functions for the superconducting fluctuations as follows
\begin{eqnarray}
&&D^R(1,2) = -i \theta(t_1-t_2)\langle \left[\Delta(1),\Delta^*(2)\right]\rangle = -i\langle \Delta_{q}(1)
\Delta_{cl}^*(2)\rangle \label{DR}\\
&&D^A(1,2) = i \theta(t_2-t_1)\langle \left[\Delta(1),\Delta^*(2)\right]\rangle = -i\langle \Delta_{cl}(1)
\Delta_{q}^*(2)\rangle \label{DA}\\
&&D^K(1,2) = -i \langle \{\Delta(1),\Delta^*(2)\}\rangle = -i\langle \Delta_{cl}(1)
\Delta_{cl}^*(2)\rangle \label{DK}
\end{eqnarray}
From Eqns~\ref{SK2E},~\ref{PiRexp2} and~\ref{PiKneq}, the equation of motion obeyed by the above Green's functions are
\begin{eqnarray}
D^K = -D^R \Pi^K D^A \label{eomDK}
\end{eqnarray}
where
\begin{eqnarray}
\left[\eta \left(\frac{\partial}{\partial t_1} -i \vec{v}_D\cdot\left(\vec{q} + 2 e \vec{E} t_1\right)\right)+
\delta + \gamma \left( \vec{q} + 2 e \vec{E} t_1\right)^2 \right]D^R(\vec{q}; t_1,t_2) = - \delta(t_1,t_2)
\label{eomDR}
\end{eqnarray}
The above equation corresponds to overdamped dynamics and may be solved easily,
\begin{eqnarray}
D^R(\vec{q};t_1,t_2) = -\theta(t_1-t_2) \frac{1}{\eta} \exp{\left(-\frac{1}{\eta}\int_{t_2}^{t_{1}}d\tau
\left[ \epsilon_q(\tau) -i \eta \vec{v}_D\cdot\left(\vec{q} + 2 e \vec{E} \tau\right)\right]\right)}
\label{DR1}
\\
D^A(q;t_1,t_2) = -\theta(t_2-t_1) \frac{1}{\eta} \exp{\left(\frac{1}{\eta}\int_{t_2}^{t_{1}}d\tau
\left[ \epsilon_q(\tau)+ i \eta \vec{v}_D\cdot\left(\vec{q} + 2 e \vec{E} \tau\right) \right]\right)}
\label{DA1}
\end{eqnarray}
where
\begin{equation}
\epsilon_q(\tau) = \delta + \gamma \left(\vec{q} + 2 e \vec{E} \tau \right)^2 \label{eq}
\end{equation}

\subsection{Self-consistent gap equation}

The self-consistent gap equation is
\begin{eqnarray}
&&\delta = \delta_0 + u_1 \langle |\Delta_{cl}|^2\rangle
\label{selfcon1}
\end{eqnarray}
where $u_1 \sim \frac{\gamma}{\eta}\nu \lambda \tau_{sc}^2$ and 
\begin{eqnarray}
\langle |\Delta_{cl}|^2\rangle = i D^K(x,t;x,t) = -i \int d2 d3 D^R(1,2) \Pi^K(2,3) D^A(3,1) \nonumber \\
= -i \int \frac{d^2q}{(2\pi)^2} \int_{-\infty}^{t} dt_1 \int_{-\infty}^t dt_2
D^R(q;t,t_1) \Pi^K(t_1,t_2) D^A(q;t_2,t)
\end{eqnarray}
In Fourier space $\Pi^K(t_1,t_2)= \int \frac{d\Omega}{2\pi}e^{-i \Omega(t_1-t_2)}\Pi^K(\Omega)$
which together with Eqns~\ref{DR1},~\ref{DA1},~\ref{PiKneq}
give
\begin{eqnarray}
&&\langle |\Delta_{cl}|^2\rangle = -i
\int \frac{d^2q}{(2\pi)^2} \int \frac{d\Omega}{2\pi}
\Pi^K(\Omega)\int_{-\infty}^{t} dt_1 \int_{-\infty}^t dt_2
\frac{1}{\eta^2}e^{-i \Omega(t_1-t_2)}
e^{-\frac{1}{\eta}\int_{t_1}^{t}d\tau_1\left[\delta + \gamma (\vec{q} + 2e \vec{E}\tau_1)^2 -i \eta
\vec{v}_D\cdot\left(\vec{q} + 2 e \vec{E} \tau_1\right)\right]}  \nonumber \\
&&e^{-\frac{1}{\eta}\int_{t_2}^{t}d\tau_2\left[\delta + \gamma (\vec{q} + 2e \vec{E}\tau_2)^2 +i \eta
\vec{v}_D\cdot\left(\vec{q} + 2 e \vec{E} \tau_2\right)\right]}
\end{eqnarray}
Changing variables to the canonical
momentum $\vec{k} = \vec{q} + 2 e \vec{E} t$, the explicit dependence on $t$ goes away, and
one obtains
\begin{eqnarray}
&&\langle |\Delta_{cl}|^2\rangle = -i
\int \frac{d^2k}{(2\pi)^2} \int \frac{d\Omega}{2\pi}
\Pi^K(\Omega)
\frac{1}{\eta^2} \int_0^{\infty} d x \int_{0}^{\infty} dy \nonumber \\
&&e^{i\Omega(x-y) + i \vec{v}_D\cdot\vec{k}(x-y)-i \vec{v}_D\cdot e\vec{E}(x^2 - y^2)}
e^{-\frac{1}{\eta}\left( \delta + \gamma k^2 \right)(x+y)}
e^{-\frac{\gamma}{3\eta}(2 e E)^2(x^3 + y^3)}  e^{\frac{\gamma}{\eta}\vec{k}\cdot 2e\vec{E}(x^2 + y^2)}
\end{eqnarray}
It is convenient to perform the momentum integrals, which gives,
\begin{eqnarray}
&&\langle |\Delta_{cl}|^2\rangle = \frac{-i}{4 \pi^2} \frac{\eta \pi}{\gamma}
\int \frac{d\Omega}{2\pi}
\Pi^K(\Omega)
\frac{1}{\eta^2} \int_0^{\infty} d x \int_{0}^{\infty} dy e^{i\Omega(x-y)-i \vec{v}_D\cdot e\vec{E}(x^2 - y^2)}
\frac{1}{x+y}
e^{-\frac{\delta}{\eta}(x+y)} \nonumber \\
&& e^{-\frac{4 \gamma}{3\eta}(e E)^2(x^3 + y^3)}
e^{\frac{\gamma}{\eta}(e E)^2 \frac{(x^2 + y^2)^2}{x+y}\left(1 + i\frac{v_D}{eE}
(\frac{\eta}{2\gamma})\frac{x-y}{x^2 + y^2} \right)^2}
\end{eqnarray}
After this the manipulations are similar to~\cite{Phillips04}.
It is convenient to change variables to $u= x + y, v= x-y$, so that $\int_0^{\infty}dx \int_0^{\infty} dy
=\frac{1}{2}\int_0^{\infty} d u \int_{-u}^{u}d v$ giving
\begin{eqnarray}
&&\langle |\Delta_{cl}|^2\rangle = \frac{-i}{4 \pi^2} \frac{\eta \pi}{\gamma}
\int \frac{d\Omega}{2\pi}
\Pi^K(\Omega)
\frac{1}{2\eta^2} \int_0^{\infty} \frac{d u}{u} \int_{-u}^{u} dv  \nonumber \\
&&e^{i\Omega v- i \vec{v}_D \cdot e\vec{E}u v}
e^{-\frac{\delta}{\eta}u}
e^{-\frac{\gamma}{3\eta}(e E)^2 u (u^2 + 3 v^2)}  e^{\frac{\gamma}{4\eta}(e E)^2 \frac{(u^2 + v^2)^2}{u}
\left(1 + \frac{i v_D \eta }{e E \gamma} \frac{v}{u^2 + v^2}\right)^2}
\end{eqnarray}

Now we approximate the expression for $\Pi^K$ in Eq.~\ref{PiKneq} as
$\Pi^K(\Omega) \simeq 2 i \eta \left[ |\Omega| +\frac{2 T_{eff}}{\pi} \theta(|\Omega| - T_{eff})\right]$. We also define dimensionless variables
$\bar{\Omega}=\Omega \eta,
u/\eta \rightarrow u, v/\eta \rightarrow v$
\begin{eqnarray}
\bar{T}_{eff}= T_{eff}\eta \label{Tdef1}\\
\bar{E}= e E \eta \sqrt{\gamma} \label{Edef1}
\end{eqnarray}
in terms of which
\begin{eqnarray}
&&\langle |\Delta_{cl}|^2\rangle = \frac{1}{2 \pi \gamma \eta}
\int_{-\infty}^{\infty} \frac{d\bar{\Omega}}{2\pi}
\left[|\bar{\Omega}| + \frac{2 \bar{T}_{eff}}{\pi} \theta(|\Omega| - T_{eff}) \right]
\int_0^{\infty} \frac{du}{u} \int_{0}^{u} dv \cos{\left(\bar{\Omega} v -\frac{\tau_{sc}}{\gamma m}\bar{E}^2 u v +
\frac{\tau_{sc}}{2 \gamma m}\bar{E}^2 \frac{v}{u}\right)} \nonumber \\
&&e^{-{\delta}u}
e^{-\frac{\bar{E}^2}{3}u (u^2 + 3 v^2)}  e^{\frac{\bar{E}^2}{4}\frac{(u^2 + v^2)^2}{u}\left(1 -
\frac{\tau_{sc}^2}{\gamma^2 m^2} \frac{v^2}{(u^2 + v^2)^2}\right)}
\end{eqnarray}
Changing variables to $v \rightarrow v/u$, one gets
\begin{eqnarray}
&&\langle |\Delta_{cl}|^2\rangle = \frac{1}{2 \pi \gamma \eta}
\int_{-\infty}^{\infty} \frac{d\bar{\Omega}}{2\pi}
\left[|\bar{\Omega}| + \frac{2 \bar{T}_{eff}}{\pi} \theta(|\Omega| - T_{eff})\right]
\int_0^{\infty} du \int_{0}^{1} dv \cos{\left(\bar{\Omega} u v - \frac{\tau_{sc}}{\gamma m}\bar{E}^2 u^2
+ \frac{\tau_{sc}}{2\gamma m}\bar{E}^2 v\right)}  \nonumber \\
&&e^{-{\delta}u}
e^{-\frac{\bar{E}^2u^3}{12}\left(1 + 6 v^2 - 3 v^4 \right) - \frac{\bar{E}^2}{4}\frac{\tau_{sc}^2}{\gamma^2 m^2}
u v^2 (1 + v^2)}\label{Dcl}
\end{eqnarray}
It is now straightforward to see the role played by current drift. This
term always arises in the combination $\frac{\tau_{sc}}{\gamma m}\bar{E}^2$. Using
Eq.~\ref{gamma}, one finds it to be ${\cal O}\left( \frac{\bar{E}^2}{\mu \tau_{sc}}\right)$.
As we shall show, the electric field scaling due to the current noise term is ${\cal O}({\bar E}^2)$
in the quantum-disordered regime.
Thus the drift gives corrections to this result by an amount which is smaller by a factor of
$\frac{1}{E_F \tau_{sc}} \ll 1$
($E_F =\mu$). Therefore in what follows we will drop the drift term from further analysis.

Substituting Eq.~\ref{Dcl} in Eq.~\ref{selfcon1}, and adding subtracting terms one gets the following
self-consistent gap equation
\begin{eqnarray}
&&\delta = \delta_0 + \frac{u_1}{2\pi \gamma \eta}\int_{-\infty}^{\infty} \frac{d\bar{\Omega}}{2\pi}
|\bar{\Omega}|
\int_0^{\infty} du \int_{0}^{1} dv \cos{\left(\bar{\Omega} u v\right)}
e^{-{\delta}u} \label{selfcon2}\\
&&+\frac{u_1}{2\pi \gamma \eta}\int_{-\infty}^{\infty} \frac{d\bar{\Omega}}{2\pi}
|\bar{\Omega}|
\int_0^{\infty} du \int_{0}^{1} dv \cos{\left(\bar{\Omega} u v\right)}
e^{-{\delta}u}\left[e^{-\frac{\bar{E}^2u^3}{12}\left(1 + 6 v^2 - 3 v^4 \right)}-
e^{-\frac{\bar{E}^2u^3}{12}} + e^{-\frac{\bar{E}^2u^3}{12}} -1  \right] \nonumber \\
&&+ \frac{u_1}{2 \pi \gamma \eta}
\int_{-\infty}^{\infty} \frac{d\bar{\Omega}}{2\pi}
\left[\frac{2 \bar{T}_{eff}}{\pi} \theta(|\Omega| - T_{eff}) \right]
\int_0^{\infty} du \int_{0}^{1} dv \cos{\left(\bar{\Omega} u v\right)}
e^{-{\delta}u}
e^{-\frac{\bar{E}^2u^3}{12}\left(1 + 6 v^2 - 3 v^4 \right)}
\end{eqnarray}
We introduce a frequency cutoff $\Lambda$ in the first term in the above equation, and perform
the frequency integral to obtain,
\begin{eqnarray}
&&\delta\left[1 +\frac{u_1}{2\pi^2 \gamma \eta} \ln \frac{1}{\delta}  \right]
= \left[\delta_0 + \frac{u_1}{2\pi^2 \gamma \eta}\left(\Lambda \frac{\pi}{2} - \delta \ln \Lambda\right) \right]
\label{selfcon3}\\
&&+\frac{u_1}{2\pi \gamma \eta}\int_{-\infty}^{\infty} \frac{d\bar{\Omega}}{2\pi}
|\bar{\Omega}|
\int_0^{\infty} du \int_{0}^{1} dv \cos{\left(\bar{\Omega} u v\right)}
e^{-{\delta}u}\left[e^{-\frac{\bar{E}^2u^3}{12}\left(1 + 6 v^2 - 3 v^4 \right)}-
e^{-\frac{\bar{E}^2u^3}{12}} + e^{-\frac{\bar{E}^2u^3}{12}} -1  \right] \nonumber \\
&&+ \frac{u_1}{2 \pi \gamma \eta}
\int_{-\infty}^{\infty} \frac{d\bar{\Omega}}{2\pi}
\left[\frac{2 \bar{T}_{eff}}{\pi} \theta(|\Omega| - T_{eff})
\right]
\int_0^{\infty} du \int_{0}^{1} dv \cos{\left(\bar{\Omega} u v\right)}
e^{-{\delta}u}
e^{-\frac{\bar{E}^2u^3}{12}\left(1 + 6 v^2 - 3 v^4 \right)}
\end{eqnarray}
The remaining frequency integrals above are performed by
introducing a cutoff $e^{- \Lambda^{-1}|\Omega|}$ in the argument, and then setting $\Lambda^{-1}=0$. For
example, one integral evaluates to
$Lt_{\Lambda^{-1}\rightarrow 0} \int_0^{\infty}\Omega \cos{\left(\Omega u v\right)} e^{-\Omega/\Lambda}
= \frac{\Lambda^{-1} - u^2 v^2}{(\Lambda^{-2} + u^2 v^2)^2} = \frac{-1}{u^2 v^2}$, while another is
$Lt_{\Lambda^{-1}\rightarrow 0} \int_0^{\infty} d\Omega \sin{(\Omega u)} e^{- \Omega/\Lambda}
= \frac{1}{u}$.

In addition, by defining,
$\delta^R = \frac{\delta_0}{{\frac{u_1}{2\pi^2 \gamma \eta}}} +
\left[\Lambda \frac{\pi}{2} - \delta \ln \Lambda \right] $ as the
renormalized distance from the QCP, and by using $\ln1/\delta \gg
1$, the self-consistent gap equation becomes,
\begin{eqnarray}
&&\delta \ln \frac{1}{\delta} = \delta^R -\int_0^{\infty} du
\int_{0}^{1} dv \frac{1}{u^2 v^2}
e^{-{\delta}u}\left[e^{-\frac{\bar{E}^2u^3}{12}\left(1 + 6 v^2 - 3
v^4 \right)}-
e^{-\frac{\bar{E}^2u^3}{12}}\right] \label{selfcon7}\\
&& +
\int_0^{\infty} du \frac{1}{u^2}
e^{-{\delta}u}\left[e^{-\frac{\bar{E}^2u^3}{12}} -1  \right]
\nonumber + \left(\frac{2 \bar{T}_{eff}}{\pi}\right)
\int_0^{\infty} du \int_{0}^{1} dv \frac{1}{u v}\sin{\left(\bar{T}_{eff} u v\right)}
e^{-{\delta}u}
e^{-\frac{\bar{E}^2u^3}{12}\left(1 + 6 v^2 - 3 v^4 \right)}
\end{eqnarray}
The first three terms on the r.h.s was derived
in~\cite{Phillips04}, whereas the last term is new and arises due
to current noise and reflects the modification of the underlying electron distribution function.

We discuss the solution of the gap equation in two regimes \newline
{\bf A. Quantum Disordered Regime, $\delta^R \gg \bar{E}$}. Here
Eq.~\ref{selfcon7} can be perturbatively expanded in powers of
$\bar{E}$ to give
\begin{equation}
\delta \simeq \frac{\delta^R}{\ln \frac{1}{\delta^R}} +
\frac{\bar{E}^2}{3 (\delta^R/\ln\frac{1}{\delta^R})^2} +
\frac{2}{\pi} \frac{\bar{T}_{eff}^2}{(\delta^R /\ln
\frac{1}{\delta^R})} \label{quandis}
\end{equation}
While the first two terms in Eq.~\ref{quandis} were derived
in~\cite{Phillips04}, the last term is the correction due to current
noise which essentially acts as an effective temperature. As
discussed before, current drift will correct this result by a factor
of ${\cal O}(1/E_F \tau_{sc})$.
\newline
{\bf B.  Quantum Critical Regime, $\delta^R \ll \bar{E}$}. Here one
may set $e^{\delta u}=1$ in Eq.~\ref{selfcon7}, which in terms of a
rescaled variable $\bar{u} = u E^{2/3}$, may be written as
\begin{eqnarray}
&&\delta \ln \frac{1}{\delta}
= \bar{E}^{2/3}\left[\int_0^{\infty} d\bar{u} \frac{1}{\bar{u}^2}
\left(e^{-\frac{\bar{u}^3}{12}} -1  \right) -\int_0^{\infty} d\bar{u} \int_{0}^{1} dv \frac{1}{\bar{u}^2 v^2}
\left(e^{-\frac{\bar{u}^3}{12}\left(1 + 6 v^2 - 3 v^4 \right)}-
e^{-\frac{\bar{u}^3}{12}}\right) \right] \label{quantcrit}\\
&&+\left(\frac{2 \bar{T}^2_{eff}}{\pi \bar{E}^{2/3}}\right)
\int_0^{\infty} d\bar{u} \int_{0}^{1} dv
e^{-\frac{\bar{u}^3}{12}\left(1 + 6 v^2 - 3 v^4 \right)}
\nonumber
\end{eqnarray}
Defining the following functions
\begin{eqnarray}
{\cal Y} = \frac{1}{3^{1/3}2^{4/3}}\Gamma\left( \frac{2}{3}\right)
\int_0^{1} dv \frac{\left[\left(1 + 6 v^2 - 3 v^4 \right)^{1/3}- (1 + v^2) \right]}{v^2}
=0.1165 \\
{\cal Y^{\prime}} = \int_0^{\infty} d u \int_0^{1} d ve^{- \frac{u^3}{12}\left(1 + 6 v^2 - 3 v^4 \right)}
=1.603
\end{eqnarray}
we find,
\begin{eqnarray}
\delta \simeq \frac{\left( 2 \bar{E}\right)^{2/3}{\cal Y}}{\ln \frac{1}{(2\bar{E})^{2/3}}}
\left[ 1 + 5.52 \frac{\bar{T}_{eff}^2}{\bar{E}^{4/3}} \right] \label{delqc}
\end{eqnarray}
Again the first term was derived in~\cite{Phillips04}, while the
second term above is the correction arising due to the modification
of the distribution function of the underlying electrons. Using the definitions Eq.~\ref{Tdef1},~\ref{Edef1}
and the expressions for $\eta$ and $\gamma$ in Eqns.~\ref{alpha},~\ref{gamma}, 
$\frac{\bar{T}_{eff}^2}{\bar{E}^{4/3}}\sim \left( T_{eff} \tau_{sc}\right)^{2/3}\ll 1$. Thus 
this term only gives rise to subleading corrections within the model presented here where
$\tau_{sc}$ is independent of the electric-field. In the conclusions we discuss the case of 
systems where $\tau_{sc}$ may have a strong electric-field dependence,
and can in particular diverge as $E \rightarrow 0$. In this case it may be
possible for the second 
term to dominate over the first. 

\subsection{Expression for the current due to superconducting fluctuations}

We now turn to the evaluation of current due to superconducting fluctuations.
The expression for the current is given by
\begin{eqnarray}
\vec{J} = \frac{\delta Z_K}{\delta \vec{A}} = \frac{2 e}{\hbar}\gamma
\int \frac{d^2q}{(2\pi)^2} \left(\vec{q} + 2 e \vec{E}t\right) iD^K(q;t,t) \label{j1}
\end{eqnarray}
Changing variables to the canonical
momentum $\vec{k} = \vec{q}+ 2 e \vec{E} t$, and using Eq.~\ref{eomDK}, we obtain the
expression
\begin{eqnarray}
\vec{J} = -i\frac{2 e}{\hbar}\gamma\int \frac{d^2k}{(2\pi)^2}\vec{k}
\int \frac{d\Omega}{2\pi}
\Pi^K(\Omega)
\frac{1}{\eta^2} \int_0^{\infty} d x \int_{0}^{\infty} dy e^{i\Omega(x-y)}
e^{-\frac{1}{\eta}\left( \delta + \gamma k^2 \right)(x+y)}
e^{-\frac{\gamma}{3\eta}(2 e E)^2(x^3 + y^3)}  e^{\frac{\gamma}{\eta}\vec{k}\cdot 2e\vec{E}(x^2 + y^2)}
\label{j2}
\end{eqnarray}
Performing the momentum integral, one gets
\begin{eqnarray}
J  = \frac{-i}{4 \pi^2} \frac{\eta \pi}{\gamma} \left( \frac{2e}{\hbar}\right)\gamma e E
\int \frac{d\Omega}{2\pi}
\Pi^K(\Omega)
\frac{1}{\eta^2} \int_0^{\infty} d x \int_{0}^{\infty} dy e^{i\Omega(x-y)} \frac{x^2 + y^2}{(x+y)^2}
e^{-\frac{\delta}{\eta}(x+y)}
e^{-\frac{4 \gamma}{3\eta}(e E)^2(x^3 + y^3)}  e^{\frac{\gamma}{\eta}(e E)^2 \frac{(x^2 + y^2)^2}{x+y}}
\label{j3}
\end{eqnarray}
As before we perform a change of variables to previously defined dimensionless
variables to obtain,
\begin{eqnarray}
J  = \frac{2 e^2}{\hbar}\frac{E}{2 \pi}
\int_{-\infty}^{\infty} \frac{d\bar{\Omega}}{2\pi}
\left[|\bar{\Omega}| + \frac{2 \bar{T}_{eff}}{\pi} \theta(|\Omega| - T_{eff})\right]
\int_0^{\infty} du \int_{0}^{1} dv \cos{\left(\bar{\Omega} u v\right)} \frac{u}{2}(1 + v^2)
e^{-{\delta}u}
e^{-\frac{\bar{E}^2u^3}{12}\left(1 + 6 v^2 - 3 v^4 \right)}
\label{j4}
\end{eqnarray}
Adding and subtracting terms in Eq~\ref{j4},
\begin{eqnarray}
&&J  =\frac{2 e^2}{\hbar}\frac{E}{2 \pi}
\int_{-\infty}^{\infty} \frac{d\bar{\Omega}}{2\pi}|\bar{\Omega}|
\int_0^{\infty} du \int_{0}^{1} dv \cos{\left(\bar{\Omega} u v\right)} \frac{u}{2}(1 + v^2)
e^{-{\delta}u}
\left[ e^{-\frac{\bar{E}^2u^3}{12}\left(1 + 6 v^2 - 3 v^4 \right)} - e^{-\frac{\bar{E}^2u^3}{12}}
+ e^{-\frac{\bar{E}^2u^3}{12}}\right]\label{j5} \\
&&+\frac{2 e^2}{\hbar}\frac{E}{2 \pi}\left(\frac{2 \bar{T}_{eff}}{\pi}\right)
\int_{-\infty}^{\infty} \frac{d\bar{\Omega}}{2\pi}
\left[\theta(|\Omega| - T_{eff})\right]
\int_0^{\infty} du \int_{0}^{1} dv \cos{\left(\bar{\Omega} u v\right)} \frac{u}{2}(1 + v^2)
e^{-{\delta}u}
e^{-\frac{\bar{E}^2u^3}{12}\left(1 + 6 v^2 - 3 v^4 \right)} \nonumber
\end{eqnarray}
One of the integrals that may be performed is $ Lt_{\Lambda^{-1}\rightarrow 0}
\int_0^{\infty} d \Omega \Omega e^{-\Omega/\Lambda}\int_0^{1} dv\cos(\Omega u v)\left(1 + v^2\right)=
\int_0^{\infty} d \Omega \Omega e^{-\Omega/\Lambda}\frac{2}{u^3\Omega^3 }
\left[u \Omega \cos(u \Omega) + (-1 + u^2 \Omega^2)\sin(u \Omega) \right]
= Lt_{\Lambda^{-1}\rightarrow 0} \frac{2}{u^3}
\left[-\frac{u}{ 1 + \Lambda^2 u^2}+ \frac{\pi/2}{\Lambda} \right] = 0 $.
Moreover using $Lt_{\Lambda^{-1}\rightarrow 0} \int_0^{\infty}\Omega \cos{\left(\Omega u v\right)} e^{-\Omega/\Lambda}
= \frac{\Lambda^{-1} - u^2 v^2}{(\Lambda^{-2} + u^2 v^2)^2} = \frac{-1}{u^2 v^2}$, the expression for the
current becomes
\begin{eqnarray}
&&J  = -\frac{e^2 E}{\hbar \pi^2}
\int_0^{\infty} du \int_{0}^{1} dv \frac{1}{u^2 v^2}\frac{u}{2}(1 + v^2)
e^{-{\delta}u}
\left[ e^{-\frac{\bar{E}^2u^3}{12}\left(1 + 6 v^2 - 3 v^4 \right)} - e^{-\frac{\bar{E}^2u^3}{12}}
\right]\label{j6} \\
&&+\frac{e^2 E}{ \hbar \pi^2}\left(\frac{2 \bar{T}_{eff}}{\pi}\right)
\int_0^{\infty} du \int_{0}^{1} dv \frac{1}{uv}\sin{\left(\bar{T}_{eff} u v\right)} \frac{u}{2}(1 + v^2)
e^{-{\delta}u}
e^{-\frac{\bar{E}^2u^3}{12}\left(1 + 6 v^2 - 3 v^4 \right)} \nonumber
\end{eqnarray}

As before we discuss the following two cases\newline {\bf A. Quantum
Disordered Regime, $\delta^R \gg \bar{E}$}. In this regime we find
\begin{equation}
J = \frac{e^2 E}{\hbar \pi^2} \left[\frac{8}{15}\frac{\bar{E}^2}{\delta^3} + \frac{4}{3\pi}
\frac{\bar{T}_{eff}^2}{\delta^2}\right]
\end{equation}
The second term above is the correction to the results
of~\cite{Phillips04} due to the effective temperature of the
nonequilibrium electrons.
\newline
{\bf B.  Quantum Critical Regime, $\delta^R \ll \bar{E}$}. Here we
obtain the result
\begin{equation}
J = \frac{e^2 E}{\hbar \pi^2}\left[ \frac{1}{6} \int_0^{1}d v \frac{(1 + v^2)}{v^2}\ln{(1 + 6 v^2 - 3 v^4)}
+ \frac{\bar{T}_{eff}^2}{\pi \bar{E}^{4/3}} \int_0^{\infty} du \int_0^1 dv
u (1 + v^2)e^{-\frac{u^3}{12}(1 + 6 v^2 - 3 v^4)}\right]
\end{equation}
Computing the above integrals we find
\begin{eqnarray}
J = \frac{0.46e^2 E}{h}\left[1 + 
 0.82\frac{\bar{T}_{eff}^2}{\bar{E}^{4/3}} \right] \label{jqc}
\end{eqnarray}
The first term is the universal conductivity found
in~\cite{Phillips04}, while the second term is the contribution due to current noise. As discussed after 
Eq.~\ref{delqc}, this correction is of ${\cal O}\left((T_{eff} \tau_{sc})^{2/3}\right)$ 
and is therefore subleading for this model of electric field independent $\tau_{sc}$.

It is instructive to see how the current due to superconducting
fluctuations in the quantum critical regime get modified for a $1d$
system. The steps in the derivation are the same except that there
is only one momentum integral in Eq.~\ref{j2}. We find
\begin{eqnarray}
J_{1d} = \frac{e^2 E \sqrt{\gamma}}{\hbar \pi \sqrt{\pi}}
\left[\bar{E}^{1/3}
\frac{2^{2/3}}{3^{1/6}}\Gamma\left(\frac{5}{6}\right)\int_0^1 dv
\frac{1 + v^2}{v^2}\left(g^{1/6}(v) -1\right) + \frac{2
\bar{T}_{eff}^2}{\pi \bar{E}}\int_0^{\infty} du \int_0^1 dv
\sqrt{u}(1 + v^2) e^{-\frac{u^3}{12}g(v)} \right] \label{j1d}
\end{eqnarray}
where $g(v) = 1 + 6 v^2 - 3 v^4 $.

Eq.~\ref{j1d} shows that unlike 2d, the response to the electric
field in the quantum critical regime is highly nonlinear, with
$J_{1d} \propto E^{4/3}$. Current noise here too gives subleading
corrections of ${\cal O}((T_{eff}\tau_{sc})^{2/3})$.

The results presented above are for the case of $\delta > 0$, {\sl
i.e,} the system is on the normal side in equilibrium. The case of
$\delta< 0$ and large electric fields so that one is on the
current/supercurrent induced disordered side can be analyzed by
employing a purely classical Ginzburg-Landau theory corresponding to
a temperature $T = T_{eff}$. The computation of the non-linear
response would follow~\cite{Mishonov02}, where using their results
one expects the fluctuation current in dimension $d$ to be, $J_{d}
\propto \frac{T_{eff}}{E^{(4-d)/3}} E $.

\section{Conclusions} \label{concl}

In summary starting from a fermionic model under external drive, we
have presented a microscopic derivation of the effect of current flow
on a superconducting order-parameter. Our microscopic treatment
reveals that current besides directly coupling to the
order-parameter also produces a noise and a drift of the
order-parameter, the origin of which is the underlying
nonequilibrium electron gas. We study the effects of these new terms on scaling near
the equilibrium quantum critical point. Scaling equations when only the direct coupling between
the order-parameter and the electric field is present was derived by  Dalidovich and
Phillips~\cite{Phillips04} in a phenomenological approach. Here we find
that current drift gives a small correction of 
${\cal O}(1/E_F\tau_{sc})$ to their result.  
Current noise on the other hand   
gives corrections that are of ${\cal O}((T_{eff}\tau_{sc})^{2/3})$ in
the quantum critical regime.  
In our model where $\tau_{sc}$ is independent of the electric
field, this correction is subdominant to the effect of the direct coupling between
the order-parameter and the electric-field. In the quantum disordered regime however
the noise and direct-coupling effects are found to be equally dominant.

One may easily imagine a scenario where noise effects dominate over direct-coupling 
effects both in the quantum-critical and quantum disordered regime. This would
occur when $\tau_{sc}\sim T_{eff}^{-p}$ where $p > 1$, a  physical situation for this
being when the dominant inelastic scattering mechanism is due to phonons.  There are
several experiments involving electric-field scaling in thin films near a superconducting transition
~\cite{Goldman,Yazdani95,Yoon06}. 
As
discussed in~\cite{Goldman}, the results of many of these experiments can be explained only when
taking into account noise effects due to a nonequilibrum electron gas. For example 
$p=2$ for electron-phonon coupled MoGe thin films, clearly making $T_{eff} \tau_{sc} \gg 1$
in these systems. 

Our derivation is valid on the normal side and outside the Ginzburg
regime. Extension of the results of this paper to the nonequilibrium
ordered side is currently in progress.

{\sl Acknowledgments}\newline A.M. thanks L. Ioffe, S. Khlebnikov, A. J. Millis, A. Polkovnikov, T. Senthil
and E. Yuzbashyan for helpful discussions. This
work was supported by NSF-DMR 0705584.

\appendix

\section{Effective equilibrium action for fluctuations about the ordered state} \label{flucord}

In order to understand the fluctuational properties on the ordered side in the absence of an applied electric field, the action will be derived for a partition function,
\begin{equation}
Z = \int {\cal D}\left[ \Delta, \Delta^*\right] 
\exp{\left(-\int d\tau d^d r \frac{|\Delta|^2}{\lambda} + Tr \ln {\cal G}^{-1}\right)}
\end{equation}
where in terms of a complex $\Delta = \Delta_0 e^{2i\theta}, \Delta^* =\Delta_0 e^{-2i\theta}$
\begin{eqnarray}
{\cal G}^{-1} = \begin{pmatrix} -\partial_{\tau} - \frac{1}{2m}\left(\frac{\vec{\nabla}}{i}- \frac{e}{c}
\vec{A}\right)^2 - \Sigma
+\mu
& \Delta_0 e^{2i\theta}
\\ \Delta_0 e^{-2i\theta} &  -\partial_{\tau} + \frac{1}{2m}\left(\frac{\vec{\nabla}}{i}+ \frac{e}{c} \vec{A}\right)^2
- \bar{\Sigma} -\mu
\end{pmatrix}
\end{eqnarray}
$\Sigma=\bar{\Sigma}$
are the self-energies due to coupling to the underlying metallic substrate with
\begin{equation}
\Sigma(\tau) = \frac{\Gamma}{\pi} P \left(\frac{1}{\tau}\right)
\end{equation}
The action may be written as an
expansion in fluctuations in the magnitude $\Delta_0$ and phase $\theta$ of the order parameter. In what follows
we will consider only fluctuations in the phase as the fluctuations in the magnitude of $\Delta$ are gapped in
the ordered phase. To this end it is convenient to introduce the unitary matrix
$U = \begin{pmatrix}e^{-i\theta}& 0 \\ 0 & e^{i\theta} \end{pmatrix}$, and transform the Green's function as
\begin{eqnarray}
&&{\cal G}^{-1}\rightarrow U {\cal G}^{-1} U^{\dagger} =\\
&&\begin{pmatrix} -\partial_{\tau}-i \phi - \frac{1}{2m}\left(\frac{\vec{\nabla}}{i}- \frac{e}{c}
\vec{\tilde{A}}\right)^2 - e^{-i\theta}\Sigma e^{i\theta}
+\mu
& \Delta_0
\\ \Delta_0  &  -\partial_{\tau} + i\phi + \frac{1}{2m}\left(\frac{\vec{\nabla}}{i}+ \frac{e}{c}
\vec{\tilde{A}}\right)^2
- e^{i\theta}\bar{\Sigma}e^{-i\theta} -\mu
\end{pmatrix}
 \\
&& ={\cal G}_0^{-1} + X_{1a} + X_{1b} + X_{2}
\end{eqnarray}
where $\phi = \partial_{\tau}\theta, \vec{\tilde{A}} = \vec{A} - \frac{c}{e}\vec{\nabla}\theta$ and we have
split the above terms as follows
\begin{eqnarray}
&&{\cal G}_0^{-1} = \begin{pmatrix} i \omega_n - \xi_k + i \Gamma sgn(\omega_n)
& \Delta_0
\\ \Delta_0  &  i\omega_n + \xi_k + i\Gamma sgn(\omega_n)\end{pmatrix} \\
&&X_{1a} = -i \sigma_3 \phi + \frac{i}{2m}\sigma_0\{\vec{\nabla},\frac{e}{c}\vec{\tilde{A}}\}_+\\
&& X_{1b} = \Sigma - e^{-i \theta \sigma_3} \Sigma e^{i \theta \sigma_3}\\
&& X_2 = -\sigma_3 \frac{e^2}{2m c^2} \vec{\tilde{A}}^2
\end{eqnarray}
Expanding to quadratic order in the fluctuations, the action for the superconductor takes
the form
\begin{eqnarray}
Z = \int {\cal D}\theta \exp{\left(-\int d\tau d^d r \left[ c_1 \left(\partial_{\tau}\theta\right)^2 + c_2
\left(\vec{\nabla} \theta - \frac{e}{c}\vec{A}\right)^2\right] + S_{\Sigma}\right)}
\label{Sord1}
\end{eqnarray}
where the first two terms above are the usual ones that arise in any superconductor with the coefficients 
changed due
to coupling to an underlying substrate. In particular,
\begin{eqnarray}
&&c_1 = -\frac{1}{2}\frac{1}{\beta L^d}Tr\left[{\cal G}_0 \sigma_3 {\cal G}_0 \sigma_3\right] \\
&&c_2 = \frac{n_s}{2m} - \frac{1}{2m^2d}\frac{1}{\beta L^d} Tr\left[p^2 {\cal G}_0 \sigma_0 {\cal G}_0 \sigma_0
\right], \, d = dimension
\end{eqnarray}
The new feature is $S_{\Sigma}$ which arises specifically due to coupling to external
normal metal reservoirs and reflects the lack of gauge invariance associated with the non-conservation
of particle number in the superconducting layer. To leading order in the fluctuation of the phase,
\begin{eqnarray}
S_{\Sigma} = Tr\left[{\cal G} X_{1b} \right] = Tr\left[\left(G(xt,xt^{\prime})
\Sigma(t^{\prime},t) + \Sigma(t,t^{\prime})\bar{G}(xt^{\prime},x t)\right)
\left(1-e^{-i\theta(x,t^{\prime})}e^{i\theta(x,t)} \right)\right]
\end{eqnarray}
Evaluating the above trace we obtain a Caldiera-Leggett type local damping,
\begin{eqnarray}
S_{\Sigma} =  g\int d^d x
\frac{1}{\beta}\sum_{m}|\Omega|_m(e^{-i\theta(x)})_{\Omega_m} \left(e^{i\theta(x)} \right)_{-\Omega_m}
\label{Sdiss1}
\end{eqnarray}
where
\begin{eqnarray}
g = 4 \Gamma \left(\frac{\nu \Gamma}{\sqrt{\Delta_0^2 + \Gamma^2}}\right)
\end{eqnarray}
Fourier transforming Eq.~\ref{Sdiss1} one gets,
\begin{equation}
S_{\Sigma} = \frac{g}{2\pi}
\int d^d x \int d\tau \int d\tau^{\prime}
\left(\frac{\theta(x,\tau)-\theta(x,\tau^{\prime})}{\tau - \tau^{\prime}}\right)^2 \label{Sdiss2}
\end{equation}

\section{Derivation of steady state single particle Green's functions} \label{steadystate}

\subsection{Derivation of retarded Green's functions} \label{relgi}

To obtain the retarded Green's function in the presence of an electric field and coupling to an
external reservoir we need to solve the Dyson equation,
\begin{equation}
\left[i\partial_{t_1} - H_0(t_1) \right]G^R_0(t_1 - t_2) = \delta(t_1 - t_2)
+ \Sigma^R G^R_0 \label{dysonR}
\end{equation}
where $H_0(t) = \sum_{\vec{k}_{\perp}\sigma}
\epsilon\left ( \vec{k}_{\perp} - \frac{e \vec{A}(t)}{ \hbar c}\right)
\psi^{\dagger}_{\vec{k}_{\perp}\sigma} \psi_{\vec{k}_{\perp}\sigma} $, $\vec{A} = - c \vec{E} t$ and
\begin{equation}
\Sigma^R(k_{\perp},\omega) = \sum_{k_z} \frac{t^2_{k_z}}{\omega - \epsilon^b_{k_z,k_{\perp}} + i\delta}
\end{equation}
For energy independent tunneling amplitude, density of states, and using the
fact that $\epsilon^b_{k_z,k_{\perp}} = \epsilon^b_{k_z} + \epsilon^b_{k_{\perp}}$ the
above expression simplifies to give an energy independent self-energy
\begin{equation}
\Sigma^R(\omega) = -i \pi t^2 \rho \int  d\epsilon^b_z\delta(\omega - \epsilon^b_z -\epsilon^b_{k_{\perp}})
= -i{\Gamma}
\end{equation}
The above implies
\begin{equation}
\Sigma^R(t_1,t_2) = -i \Gamma \delta(t_1 - t_2)
\label{SR}
\end{equation}
Substituting the above in Eq.~\ref{dysonR}, it is straightforward to show that
the retarded Green's function in the presence of an electric field and coupling to leads is:
\begin{equation}
G^R_0(\vec{k}, \tau) = -i \theta(\tau) e^{-\frac{i}{\hbar}\int_{-\frac{\tau}{2}}^{\frac{\tau}{2}}
dx \epsilon(\vec{k} + \frac{e}{\hbar} \vec{E} x)}
e^{-\Gamma\tau} \label{grsol}
\end{equation}
where $\tau = t_1 - t_2$ and $ k = p + e E T$, (where we set $\hbar = 1$).
The above time integral in the argument may be performed to obtain the following series expansion
\begin{equation}
G^R_0(\vec{k}, \tau) = -i \theta(\tau) e^{-i \epsilon_k \tau -\frac{i\tau^3}{24}
\left(e \vec{E}\cdot \frac{\partial}{\partial \vec{k}} \right)^2 \epsilon_k + \ldots}
e^{-\Gamma\tau} \label{grsollin}
\end{equation}
Now we define
\begin{eqnarray}
T_{eff} = e E v_F \tau_{sc}\\
\tau_{sc}^{-1} = 2\Gamma
\end{eqnarray}
and
$E_F = v_F/a$ with $a$ being the lattice spacing.
Then, the second term in the argument of the exponent in Eq.~\ref{grsollin} is
$ (eE)^2(\tau_{sc}^3\partial^2\varepsilon_k/(\partial k^2)) =\frac{\partial^2\epsilon/a^2\partial k^2}
{E_F}\frac{ (T_{eff}\tau_{sc})^2}{E_F\tau_{sc}} \ll 1$ and therefore may be neglected. A similar argument
applies to the higher order terms.

Thus, we may approximate the retarded Green's function by its value in the absence of an electric field,
\begin{equation}
G^R_0(\vec{k}, \tau) = -i \theta(\tau) e^{-i \epsilon_k \tau}
e^{-\Gamma\tau} \label{grsollin1}
\end{equation}
provided $k$ is chosen to be the canonical momentum.

\subsection{Derivation of steady state Keldysh Green's function} \label{kelgi}

The equation of motion obeyed by the Keldysh Green's function is
\begin{eqnarray}
\left(i{\overrightarrow\partial}_{t_1} -H_0 \right)G^K_0(t_1,t_2) = 1 + \Sigma^R G^K_0 + \Sigma^K G^A_0 \label{eom1}\\
G^K_0(t_1,t_2)\left(-i {\overleftarrow {\partial}}_{t_2} - H_0 \right) = 1 + G^R_0 \Sigma^K + G^K_0 \Sigma^A
\label{eom2}
\end{eqnarray}
Taking the difference between the equations~\ref{eom1} and ~\ref{eom2},
one obtains
\begin{equation}
\left(i\partial_{t_1} + i\partial_{t_2}\right)G^K_0(t_1,t_2) - \epsilon(t_1) G^K_0(t_1,t_2) +
\epsilon(t_2) G^K_0(t_1,t_2) = \Sigma^R G^K_0 + \Sigma^K G^A_0  -G^R_0 \Sigma^K - G^K_0 \Sigma^A
\label{ke1}
\end{equation}
The solution for $G^K_0$ may be obtained by using the ansatz
\begin{equation}
G^K_0 = G^R_0 f_K - f_K G^A_0
\label{anz}
\end{equation}
where $1-2f = f_K$, with $f$ the generalized distribution function. The equation of motion for
$f^K$ is
\begin{eqnarray}
&&i \frac{\partial f_K}{\partial t_1} + i \frac{\partial f_K}{\partial t_2}-\epsilon_{p-\frac{e}{\hbar c}A(t_1)}f_K
+ \epsilon_{p-\frac{e}{\hbar c}A(t_2)} f^K \nonumber - \Sigma^R \cdot f_K + f_K \cdot \Sigma^A
+ \Sigma^K=0
\label{eomf}
\end{eqnarray}
$\Sigma^R - \Sigma^A =-\frac{i}{\tau_{sc}}$ and $\Sigma^K = (\Sigma^R - \Sigma^A)(1-2g)$,
$g$ being the distribution function of the substrate.
Fourier transforming Eq.~\ref{eomf} with respect to the relative time $\tau=t_1-t_2$, changing variables
to the canonical momentum $\vec{k} = \vec{p}+ e \vec{E}T$ and expanding in $E$
one finds that  the distribution function at
steady state obeys,
\begin{eqnarray}
&&e\vec{E}\cdot \frac{\partial f}{\partial \vec{k}} + \frac{\partial f}{\partial \omega} \left(e\vec{E}\cdot\frac{\partial \epsilon_k}{\partial
\vec{k}}\right)
 + \frac{1}{24}\frac{\partial^3 f}{\partial \omega^3}\left(e\vec{E}\cdot \frac{\partial}{\partial \vec{k}}\right)^3
\epsilon_k
\ldots =\frac{1}{\tau_{sc}}\left[-f + g \right]
\label{eomf2}
\end{eqnarray}
The usual quasiclassical arguments imply that the first term in Eq \ref{eomf2} is negligible
while in the  the weak field limit  the third term may be dropped.
With these simplifications we find
\begin{equation}
f=f^s + f^a \label{fs1}
\end{equation}
where
\begin{eqnarray}
f^s_{k,x}&=& \theta(-x) + \frac{sign(x)}{2}
e^{- \frac{|x|}{\sqrt{\left(e\vec{E}\cdot \vec{v}_k \tau_{sc}\right)^2}}}
\label{fssolph}\\
f^a_{k,x} &=&\frac{\left(e \vec{E} \cdot \vec{v}_k \tau_{sc}\right)}{2\sqrt{\left(e\vec{E}\cdot \vec{v}_k \tau_{sc} \right)^2}}
e^{- \frac{|x|}{\sqrt{\left(e\vec{E}\cdot
\vec{v}_k \tau_{sc}\right)^2}}}
\label{fasolph}
\end{eqnarray}
where $x = \omega - \mu$ and $v_k=\partial \varepsilon_k/\partial k$. Substitution of Eqs
\ref{fssolph}, \ref{fasolph} into Eq \ref{eomf2} then shows that the neglect of the third term
in Eq \ref{eomf2} is justified when the coupling of the layer to the substrate is sufficiently weak
$( \frac{\partial^3 \epsilon_k}{E_Fa^3 \partial k^3}\ll \left(E_F \tau_{sc}\right)^2)$  while the first term
is negligible in the weak field limit $T_{eff} \ll E_F^2/(\frac{\partial^2 \epsilon_k}{a^2\partial k^2})$.

\section{Evaluation of the polarization bubbles when $E \neq 0$} \label{POL}

The retarded and Keldysh polarization bubbles may be expressed as an expansion in
$\left(\vec{q} + 2 e \vec{E} T\right)^2$ and $\Omega$ as shown in Eq.~\ref{PiRexp},~\ref{PiKexp}. In particular,
\begin{eqnarray}
&&\tilde{\Pi}^R(\vec{q}+ 2 \vec{E}T=0,\Omega) = \\
&&i \sum_{\vec{k}} \int \frac{d\omega}{2\pi} \left[G_{0R}(\vec{k},\omega+ \Omega)
G_{0K}(-\vec{k},-\omega)+ G_{0K}(\vec{k},\omega+ \Omega) G_{0R}(-\vec{k},-\omega)\right]\nonumber \\
&&\Pi^K(\vec{q}+ 2 \vec{E}T=0,\Omega) = \\
&&i \sum_{\vec{k}} \int \frac{d\omega}{2\pi} \left[G_{0K}(\vec{k},\omega+ \Omega)
G_{0K}(-\vec{k},-\omega)+ G_{0R}(\vec{k},\omega+ \Omega) G_{0R}(-\vec{k},-\omega) 
+ G_{0A}(\vec{k},\omega+ \Omega) G_{0A}(-\vec{k},-\omega) \right]
\nonumber
\end{eqnarray}

Using Eq.~\ref{grsollin1},~\ref{anz},Eq.~\ref{fs1},Eq.~\ref{fssolph}~\ref{fasolph} we find
\begin{eqnarray}
{\sl Im} \left[\tilde{\Pi}^R(\Omega)\right] = \frac{i\Omega}{2} \sum_{\vec{p}} 
\int \frac{d\omega}{2\pi} \frac{(-2i\Gamma)^2}
{(\left(\omega - \xi_p\right)^2 + \Gamma^2)(\left(\omega + \xi_p\right)^2 + \Gamma^2)}
\frac{1}{\sqrt{\left( e \vec{E}\cdot \vec{v}_F \tau_{sc}\right)^2}} e^{-\frac{|\omega|}
{\sqrt{\left( e \vec{E}\cdot \vec{v}_F \tau_{sc}\right)^2}}}
\end{eqnarray}
where $2\Gamma = \tau_{sc}^{-1}$.
For $e E v_{F}\tau_{sc} < 1/\tau_{sc}$, the above expression simplifies to
\begin{equation}
{\sl Im} \left[\Pi^R(\Omega)\right] = -2i \Omega \nu \tau_{sc}
\end{equation}
In the same way, one finds
\begin{eqnarray}
\Pi^K(\Omega) = -4i\nu\tau_{sc}\left[|\Omega| + T_{eff} \int_{-\pi}^{\pi}
\frac{d\phi}{2\pi} |\cos\phi| e^{-\frac{|\Omega|}{T_{eff}|\cos\phi|}}
\right]
\label{PiKapp}
\end{eqnarray}
It is also instructive to derive the expressions for the
polarization bubbles in 1d. While $\Pi^{R,A}$ have the same
structure as in $2d$, the noise $\Pi^K$ has the form,
\begin{eqnarray}
\Pi^K_{1d}(\Omega) = -4i\nu\tau_{sc}\left[|\Omega| + T_{eff}
e^{-\frac{|\Omega|}{T_{eff}}}\right] \label{PiK1d}
\end{eqnarray}
Note that the above expressions were derived and used to study the
effect of current flow on magnetic fluctuations in~\cite{Mitra08a}.

In order to evaluate the coefficient of expansion in powers of $q + 2 e E T$
in Eq.~\ref{PiRexp2} we first write the expression for
the polarization bubble in momentum-time space
\begin{eqnarray}
&&{\Pi}^R(\vec{q}+ 2 \vec{E}T,t_1-t_2) = \nonumber \\
&&i \sum_k \left[G_{0R}(\vec{k} + \vec{q} + 2 e \vec{E} T,t_1-t_2)
G_{0K}(-\vec{k},t_1-t_2)+ G_{0K}(\vec{k} + \vec{q} + 2 e \vec{E} T,t_1-t_2) G_{0R}(-\vec{k},t_1-t_2)\right]
\end{eqnarray}
For convenience, we shift variables so that $q + 2 e E T$ appears in the argument of the retarded
functions so that
\begin{eqnarray}
&&{\Pi}^R(\vec{q}+ 2 \vec{E}T,t_1-t_2) = \nonumber \\
&&i \sum_{\vec{k}} \left[G_{0R}(\vec{k} + \vec{q} + 2 e \vec{E} T,t_1-t_2)
G_{0K}(-\vec{k},t_1-t_2)+ G_{0K}(\vec{k},t_1-t_2) G_{0R}(-\vec{k} + \vec{q} + 2 e \vec{E} T,t_1-t_2)\right] \label{PiR3}
\end{eqnarray}
Using Eq.~\ref{grsol}, the retarded Green's function can be expanded in a power series in $Q =q + 2 e E T$ as
follows,
\begin{eqnarray}
G_{0R}(\vec{k}+ \vec{Q},\tau) = 
G_{0R}(\vec{k},\tau) \left[1 -i \frac{\vec{k}\cdot\vec{Q}}{m}\tau-i \frac{Q^2}{2m} \tau - \frac{1}{2}
\left(\frac{\vec{Q}\cdot \vec{k}}{m}\right)^2 \tau^2 + \ldots \right] \label{grqexp}
\end{eqnarray}
In the above we assume quadratic dispersion. Note that in equilibrium,
the linear in $Q$ term does not survive the angle integration.
On the other hand, a non-zero current picks
a preferred direction so that this term for our case will no longer be zero. As we shall show, this
term will give rise to current-drift.

Fourier transforming Eq.~\ref{grqexp} with respect to $\tau$ we get
\begin{eqnarray}
G_{0R}(k + Q, \omega) = \left[1 - \frac{\vec{k}\cdot \vec{Q}}{m}\frac{\partial}{\partial \omega}-\frac{Q^2}{2m} \frac{\partial}{\partial \omega}
 + \frac{1}{2}
\left(\frac{\vec{Q}\cdot \vec{k}}{m}\right)^2 \frac{\partial^2}{\partial \omega^2} \right]G_{0R}(k,\omega)
\label{grqexp1}
\end{eqnarray}
Thus Eq.~\ref{grqexp1} and ~\ref{PiR3} lead to the following for the particle-hole symmetric case,
\begin{eqnarray}
\lambda\left(\Pi^R(Q,0) - \Pi^R(0,0)\right) &=&
\gamma Q^2 -\frac{i \lambda \nu}{2\Gamma} \left( \frac{e\vec{E}\cdot \vec{Q}}{m}\right)
\tau_{sc}
\end{eqnarray}
where
\begin{equation}
\gamma = - \frac{2\lambda\nu}{\pi} \frac{\mu}{m} \int d\xi \int d\omega sgn(\omega)
\left(\frac{\Gamma}{(\omega + \xi_k)^2 + \Gamma^2}\right)
\left[ \frac{(\omega - \xi_k)^3 - 3 \Gamma^2 (\omega - \xi_k)}
{((\omega - \xi_k)^2 + \Gamma^2)^3}\right] = \frac{\lambda \nu \mu}{4 m \Gamma^2}\label{gammadef}
\end{equation}

\end{document}